\documentstyle[aps,epsf]{revtex}
\begin{document}
\draft
\author{Anne-Marie Dar\'{e}, Y.M.Vilk, and A.-M.S. Tremblay}
\title{Crossover from two- to three-dimensional critical behavior for nearly
antiferromagnetic itinerant electrons.}
\address{D\'{e}partement de physique and Centre de recherche en physique du solide.\\
Universit\'{e} de Sherbrooke, Sherbrooke, Qu\'{e}bec, Canada J1K 2R1}
\date{17 January 1996, cond-mat/9602083 }
\maketitle

\begin{abstract}
The crossover from two- to three-dimensional critical behavior of nearly
antiferromagnetic itinerant electrons is studied in a regime where the
inter-plane single-particle motion of electrons is quantum-mechanically
incoherent because of thermal fluctuations. This is a relevant regime for
very anisotropic materials like the cuprates. The problem is studied within
the Two-Particle Self-Consistent approach (TPSC), that has been previously
shown to give a quantitative description of Monte Carlo data for the Hubbard
model. It is shown that TPSC belongs to the $n\rightarrow \infty $ limit of
the $O\left( n\right) $ universality class. However, contrary to the usual
approaches, cutoffs appear naturally in the microscopic TPSC theory so that
parameter-free calculations can be done for Hubbard models with arbitrary
band structure. A general discussion of universality in the
renormalized-classical crossover from $d=2$ to $d=3$ is also given.
\end{abstract}


\section{Introduction}

A simple model of itinerant antiferromagnets is provided by electrons on a
lattice with short-range repulsion. In the low temperature phase, the system
is in a Spin Density Wave (SDW) state. In three dimensions, above the
transition temperature, the electrons form a so-called {\it nearly
antiferromagnetic Fermi liquid}. Traditional mean-field techniques for
studying SDW instabilities of Fermi liquids fail completely in low
dimension. In two dimensions for example, the Random Phase Approximation
(RPA) predicts finite temperature antiferromagnetic transitions while this
is forbidden by the Mermin-Wagner theorem. Nevertheless, one can study
universal critical behavior using various forms of renormalization group
treatments appropriate either for the strong\cite{Chakravarty}\cite{Sachdev}%
\cite{Chubukov} or the weak-coupling limits\cite{Hertz}\cite{Millis}. The
self-consistent-renormalized approach of Moryia\cite{moriya} also satisfies
the Mermin-Wagner theorem in two dimensions. Since cutoff-dependent scales
are left undetermined by all these approaches they must be found by other
methods. For example, in the strong-coupling limit, the spin-stiffness
constant of the non-linear $\sigma -$model must be determined from Monte
Carlo simulations. In the weak-coupling case however, Monte Carlo
simulations are limited to very small systems, of order $10\times 10$ that
do not allow one to study much of the critical regime.

Recently, the Two-Particle Self-Consistent approach\cite{Vilk} was developed
to obtain from a microscopic model a {\it quantitative} description of
itinerant electrons not only far from phase transitions, but also in the
critical regime. It was shown\cite{Vilk} that in this approach the
Mermin-Wagner theorem is satisfied and that, away from the critical regime,
the approach gives quantitative agreement with Monte Carlo simulations of
the nearest-neighbor\cite{Vilk} and next-nearest neighbor\cite{Veilleux}
Hubbard model in two dimensions. Quantitative agreement is also obtained as
one enters the narrow critical regime accessible in Monte Carlo simulations.
The approach is restricted to the one-band Hubbard model with on-site
interaction, but is valid for arbitrary dispersion relation. The TPSC
approach also allows one to study the case where the instability of the
itinerant electrons is at an incommensurate wave-vector, but in this paper
we restrict ourselves to the case where the order is at the
antiferromagnetic wave vector. The self-consistent-renormalized approach of
Moryia\cite{moriya} cannot deal with the incommensurate case without {\it a
priori }information. Even though it has the same critical behavior as the
TPSC approach it does not allow one to obtain quantitative parameter-free
results from a microscopic Hamiltonian.

We first show in full generality that the TPSC approach gives the leading
term of the critical behavior in a $1/n$ expansion. In other words, it gives
the $n\rightarrow \infty $ limit of the $O\left( n\right) $ model where $n=3$
is the physically correct (Heisenberg) limit. It will be apparent that there
is no arbitrariness in cutoff so that, given a microscopic Hubbard model, no
parameter is left undetermined. One can go with the same theory from the
non-critical to the critical regime.

We then show that the previously studied two-dimensional critical regimes,
namely quantum-critical\cite{Sachdev} and renormalized classical\cite
{Chakravarty} are reproduced here to leading order in $1/n$. In the quantum
critical regime, one usually distinguishes two cases\cite{Sachdev}: Model A,
where the paramagnetic Fermi surface does not intersect the magnetic
Brillouin zone, and Model B where it does. This distinction is important in
the quantum critical regime because it changes the dynamical critical
exponent. In this paper, we also give results on Model C, the case of
perfect nesting. In this case, the microscopic approach shows that
modifications to frequency-independent thermodynamic properties can arise.
In particular, in the two-dimensional perfect-nesting case the usual
exponential dependence of correlation length on temperature $\exp \left(
cst/T\right) $ can be modified to be roughly $\exp \left( cst/T^3\right) $
in some temperature region of the renormalized classical regime.

Then we study the renormalized-classical crossover from $d=2$ to $d=3$ in
the highly anisotropic case of weakly coupled planes.\cite{Konno} The
general theory of such crossover is given in Appendix D, along with a
discussion of universal crossover functions. In the main text it is shown
that in the highly anisotropic case the crossover can occur in a rather
unusual regime, namely $t_{\Vert }\ll k_BT_N\ll t_{\bot }$ where $t_{\Vert
}\left( t_{\bot }\right) $ is the inter (intra) plane hopping integral and $%
T_N$ is the three-dimensional N\'{e}el temperature. This regime is unusual
because even though one is dealing with an itinerant fermion system, the
inequality $t_{\Vert }\ll k_BT_N$ means that the smallest fermionic
Matsubara frequency is larger than the dispersion in the parallel direction,
making the three-dimensional band structure irrelevant for one-particle
properties. In the language of Refs.\cite{BourbonnaisCaron}\cite{Boies},
there is ``no coherent band motion'' in the parallel direction. Physically,
the extent of the thermal de Broglie wave packet in the direction
perpendicular to the planes is smaller than the distance between planes, a
situation that does not occur in a usual Fermi liquid since in the isotropic
case the inequality $k_BT\ll E_F$ implies that the thermal de Broglie
wavelength is much larger than the lattice spacing. Another way of
describing this $t_{\Vert }\ll k_BT_N\ll t_{\bot }$ situation is to say that
the itinerant electrons become unstable at the two-particle level while
their motion in the third direction is still quasi-classical, or quantum
incoherent, at the single-particle level because of thermal fluctuations. In
the more usual situation, coherence at the one-particle level is established
before the phase transition, namely $k_BT_N\ll t_{\Vert }\ll t_{\bot }$.
These two regimes have been extensively discussed in the $d=1$ to $d=3$
crossover of Luttinger liquids by Bourbonnais and Caron\cite
{BourbonnaisCaron}\cite{Boies}.

The above single-particle incoherent regime $t_{\Vert }\ll k_BT_N\ll t_{\bot
}$ is likely to be the relevant one for high-temperature superconductors.
While the parent insulating compound $La_2CuO_4$ has been extensively
studied in the strong coupling limit, this type of compound is expected to
be in an intermediate-coupling regime. Hence, it is legitimate to approach
the problem not only from the strong-coupling limit\cite{Keimer} but also
from the weak-coupling side, especially with the TPSC approach where all
cutoffs are determined by the microscopic model. This problem is commented
on at the end of the paper. More detailed quantitative comparisons with
experiment will appear later.

\section{Two-Particle Self-Consistent approach}

We start from the Hubbard model,

\begin{equation}
H=-\sum_{<ij>\sigma }t_{i,j}\left( c_{i\sigma }^{\dagger }c_{j\sigma
}+c_{j\sigma }^{\dagger }c_{i\sigma }\right) +U\sum_in_{i\uparrow
}n_{i\downarrow \,\,\,\,}.  \label{Hubbard}
\end{equation}
In this expression, the operator $c_{i\sigma }$ destroys an electron of spin 
$\sigma $ at site $i$. Its adjoint $c_{i\sigma }^{\dagger }$ creates an
electron. The symmetric hopping matrix $t_{i,j}$ determines the band
structure. Double occupation of a site costs an energy $U$ due to the
screened Coulomb interaction. In the present section, the hopping parameters
need not be specified. We work in units where $k_B=1$, and $\hbar =1$. As an
example that occurs later, the dispersion relation in the $d$-dimensional
nearest-neighbor model when the lattice spacing is $a$ is given by 
\begin{equation}
\epsilon _{{\bf k}}=-2t\sum_{i=1}^d\left( \cos k_ia\right) .
\end{equation}
The nearest-neighbor quasi-two dimensional case will be another case of
interest later, 
\begin{equation}
\epsilon _{{\bf k}}=-2t_{\bot }\left( \cos k_xa_{\bot }+\cos k_ya_{\bot
}\right) -2t_{\Vert }\cos k_za_{\Vert }.
\end{equation}

The TPSC approach\cite{Vilk},\cite{Vilk2} can be summarized as follows. One
approximates spin and charge susceptibilities $\chi _{sp}$, $\chi _{ch}$ by
RPA-like forms but with two different effective interactions $U_{sp}$ and $%
U_{ch}$ which are then determined self-consistently. Although the
susceptibilities have an RPA functional form, the physical properties of the
theory are very different from RPA\ because of the self-consistency
conditions on $U_{sp}$ and $U_{ch}$. The necessity to have two different
effective interactions for spin and for charge is dictated by the Pauli
exclusion principle $\langle n_\sigma ^2\rangle =\langle n_\sigma \rangle $
which implies that both $\chi _{sp}$ and $\chi _{ch}$ are related to only
one local pair correlation function $\langle n_{\uparrow }n_{\downarrow
}\rangle $. Indeed, using the fluctuation-dissipation theorem in Matsubara
formalism we have the exact sum rules, 
\begin{equation}
\langle n_{\uparrow }^2\rangle +\langle n_{\downarrow }^2\rangle +2\langle
n_{\uparrow }n_{\downarrow }\rangle -n^2=\frac 1{\beta N}\sum_{\widetilde{q}%
}\chi _{ch}(\widetilde{q})
\end{equation}
and 
\begin{equation}
\langle n_{\uparrow }^2\rangle +\langle n_{\downarrow }^2\rangle -2\langle
n_{\uparrow }n_{\downarrow }\rangle =\frac 1{\beta N}\sum_{\widetilde{q}%
}\chi _{sp}(\widetilde{q})
\end{equation}
where $\beta \equiv 1/T$, $n=\langle n_{\uparrow }\rangle +\langle
n_{\downarrow }\rangle $, $\widetilde{q}=({\bf q},iq_n)$ with ${\bf q}$ the
wave vectors of an $N$ site lattice, and with $iq_n=2\pi inT$ the bosonic
Matsubara frequencies. The Pauli principle $\langle n_\sigma ^2\rangle
=\langle n_\sigma \rangle $ applied to the left-hand side of both equations
with our RPA-like forms for $\chi _{sp}$, $\chi _{ch}$ on the right-hand
side lead to 
\begin{equation}
n+2\langle n_{\uparrow }n_{\downarrow }\rangle -n^2=\frac 1{\beta N}\sum_{%
\widetilde{q}}\frac{\chi _0(\widetilde{q})}{1+\frac 12U_{ch}\chi _0(%
\widetilde{q})},  \label{sumCharge}
\end{equation}
\begin{equation}
n-2\langle n_{\uparrow }n_{\downarrow }\rangle =\frac 1{\beta N}\sum_{%
\widetilde{q}}\frac{\chi _0(\widetilde{q})}{1-\frac 12U_{sp}\chi _0(%
\widetilde{q})},  \label{sumSpin}
\end{equation}
with $\chi _0(\widetilde{q})$ the susceptibility for non-interacting
electrons.

If $\langle n_{\uparrow }n_{\downarrow }\rangle $ is known, $U_{sp}$ and $%
U_{ch}$ are determined from the above equations. This key quantity $\langle
n_{\uparrow }n_{\downarrow }\rangle $ can be obtained from Monte Carlo
simulations or by other means. However, it may be also be obtained
self-consistently\cite{Vilk} by adding to the above set of equations the
relation 
\begin{equation}
U_{sp}=g_{\uparrow \downarrow }(0)\,U\quad ;\quad g_{\uparrow \downarrow
}(0)\equiv \frac{\langle n_{\uparrow }n_{\downarrow }\rangle }{\langle
n_{\downarrow }\rangle \langle n_{\uparrow }\rangle }.  \label{Usp}
\end{equation}
Eqs.(\ref{sumSpin}) and (\ref{Usp}) define a set of self-consistent
equations for $U_{sp}$ that involve only two-particle quantities. We call
this approach Two-Particle Self-Consistent to contrast it with other
conserving approximations like Hartree-Fock or FLEX\cite{FLEX} that are
self-consistent at the one-particle level, but not at the two-particle
level. The above procedure\cite{Vilk} reproduces both Kanamori-Brueckner
screening as well as the effect of Mermin-Wagner thermal fluctuations,
giving a phase transition only at zero-temperature in two dimensions, as
discussed in the following section. Quantitative agreement with Monte Carlo
simulations on the nearest-neighbor\cite{Vilk} and next-nearest-neighbor
models \cite{Veilleux} is obtained\cite{Vilk} for all fillings and
temperatures in the weak to intermediate coupling regime $U<8t$.

We emphasize that deep in the critical regime, the {\it ansatz} Eq.(\ref{Usp}%
) fails in the sense that $g_{\uparrow \downarrow }(0)$ eventually reaches
zero at $T=0$ in the nearest-neighbor Hubbard model at half-filling while
there is no reason to believe that this really happens. The physically
appropriate choice in the renormalized classical regime described below, is
to keep the value of $g_{\uparrow \downarrow }(0)$ fixed at its
crossover-temperature value. In the numerical calculations also described
below, we are never far enough from $T_X$ to have to worry about this. The
value of $g_{\uparrow \downarrow }(0)$ is the one that is determined
self-consistently.

\section{Critical behavior of the TPSC approach in arbitrary dimension}

In this section we discuss the critical behavior of the TPSC approach in
arbitrary dimension for hypercubic systems. It is convenient to set the
lattice spacing to unity.

As one approaches a phase transition, one enters the {\it renormalized
classical }regime,\cite{Chakravarty} where classical thermal fluctuations
dominate. In this case, the universality class for {\it static} properties
is fully determined by two exponents. Dynamics must also be considered so
that one introduces a dynamical critical exponent.

We consider the case where the transition is at the antiferromagnetic wave
vector ${\bf Q}_d$ in $d$ dimensions: ${\bf Q}_2{\bf =}\left( \pi ,\pi
\right) ,$ ${\bf Q}_3{\bf =}\left( \pi ,\pi ,\pi \right) $ etc. Since ${\bf Q%
}_d$ is at the corner of the Brillouin zone, the spin susceptibility $\chi
_0\left( {\bf Q}_d\right) $ is always, by symmetry, an extremum. This
extremum is the absolute maximum at half-filling not only in the
nearest-neighbor hopping model, but also in more general models with
next-nearest-neighbor hopping\cite{Veilleux}\cite{Benard}. The
nearest-neighbor model is discussed in more details at the end of this
section. It has some special features resulting from the additional nesting
symmetry. In the two-dimensional case, we also comment on peculiarities of
nesting and on quantum-critical behavior\cite{Chubukov}\cite{Sachdev}.

\subsection{Renormalized classical regime.}

As one decreases the temperature sufficiently close to the phase transition,
there appears a small energy scale $\delta U$ that measures the proximity to
the phase transition as determined by the Stoner criterion. This scale is
defined more precisely in Eq.(\ref{DeltaU}). The key physical point is that
this energy scale is the smallest. In particular, it is smaller than the
temperature 
\begin{equation}
\delta U\ll T
\end{equation}
so that the zero Matsubara frequency representing classical behavior
dominates all others. The self-consistency conditions Eqs.(\ref{sumSpin})(%
\ref{Usp}) then lead to a strong temperature dependence of $\delta U$. This
is the renormalized-classical regime\cite{Chakravarty}. In this regime, the
antiferromagnetic correlation length $\xi $ becomes so large that\cite{Vilk2}
\begin{equation}
\xi \gg \xi _{th}
\end{equation}
where 
\begin{equation}
\xi _{th}\equiv \frac{\left\langle v_F\right\rangle }{\pi T}
\end{equation}
is the single-particle thermal de Broglie wavelength and $\left\langle
v_F\right\rangle $ is the Fermi velocity averaged over the Fermi surface.
This provides a partial justification for the usual procedure\cite{Millis}%
\cite{Hertz} that eliminates completely the Fermionic variables and
describes the system in terms of collective Bosonic variables, as is usually
done in Hubbard-Stratonovich types of approaches.\cite{Millis}\cite{Hertz}

We first show that when most of the temperature dependence of the
susceptibility comes from the temperature dependence of $\delta U$, the
RPA-like form that we have implies that {\it in any dimension }the dynamical
exponent is $z=2$ while the classical exponent $\gamma /\nu =2-\eta $ takes
the value $\gamma /\nu =2.$ The other classical exponent $\nu $ is
determined from the self-consistency condition Eq.(\ref{sumSpin}). We show
that the corresponding universality class is the same as the $n\rightarrow
\infty $ limit of the $O\left( n\right) $ classical model. This universality
class is known in turn to be the same as that of the spherical model.\cite
{Stanley} We conclude this discussion with the lower critical dimension $d=2$%
. There the exponent $\nu $ cannot strictly be defined since, as was shown
before\cite{Vilk}, the correlation length diverges exponentially at zero
temperature instead of diverging as a power law at finite temperature. This
behavior is also the one expected from the $n\rightarrow \infty $ model,
although nesting leads to different temperature dependences that are
explained further.

\subsubsection{Exponents $\gamma /\nu $ and $z$ in arbitrary dimension}

The antiferromagnetic transition is characterized by the appearance of a
small energy scale, or equivalently a large correlation length, in the
retarded spin susceptibility 
\begin{equation}
\chi _{sp}^R({\bf q,}\omega )=\frac{\chi _0^R({\bf q,}\omega )}{1-\frac 12%
U_{sp}\chi _0^R({\bf q,}\omega )}.  \label{GeneralRPA}
\end{equation}
The small energy scale is set by 
\begin{equation}
\delta U=U_{mf,c}-U_{sp}  \label{DeltaU}
\end{equation}
where the temperature-dependent ''mean-field critical'' interaction 
\begin{equation}
U_{mf,c}\equiv 2/\chi _0\left( {\bf Q}_d,0\right)
\end{equation}
is the temperature-dependent value of $U_{sp}$ at which a phase transition
would occur according to mean-field theory. In the vicinity of this point
the small energy scale $\delta U$ allows us to approximate $\chi _{sp}^R(%
{\bf q,}\omega )$ by expanding the denominator near ${\bf q\approx Q}_d$ and 
$\omega \approx 0$ to obtain,

\begin{equation}
\chi _{sp}^R({\bf q+Q}_d{\bf ,}\omega )\approx \xi ^2\frac 2{U_{sp}\xi _0^2}%
\left[ \frac 1{1+{\bf q}^2\xi ^2-i\omega \xi ^2/D}\right]  \label{chiRPA}
\end{equation}
where the antiferromagnetic correlation length is defined by 
\begin{equation}
\xi \equiv \xi _0\left( \frac{U_{sp}}{\delta U}\right) ^{1/2}  \label{ksi}
\end{equation}
with the microscopic length scale set by 
\begin{equation}
\xi _0^2\equiv \frac{-1}{2\chi _0\left( {\bf Q}_d\right) }\left. \frac{%
\partial ^2\chi _0\left( {\bf q,}0\right) }{\partial q_x^2}\right| _{{\bf q=Q%
}_d}.  \label{ksi02}
\end{equation}
The microscopic diffusion constant $D$ is defined on the other hand by 
\begin{equation}
\frac 1D\equiv \frac{\tau _0}{\xi _0^2}
\end{equation}
where the microscopic relaxation time is, 
\begin{equation}
\tau _0=\frac 1{\chi _0\left( {\bf Q}_d\right) }\left. \frac{\partial \chi
_0^R\left( {\bf Q}_d{\bf ,}\omega \right) }{\partial i\omega }\right|
_{\omega =0}.  \label{Gamma0}
\end{equation}
This relaxation-time is non-zero in both models $B$ and $C$ where the Fermi
surface intersects the magnetic Brillouin zone.

In the presence of a large correlation length $\xi $ the scaling $q\sim \xi
^{-1}$ and $\omega \sim \xi ^{-2}$ justifies the neglect of higher-order
terms in the expansion Eq.(\ref{chiRPA}). Comparing the approximate form Eq.(%
\ref{chiRPA}) with the general scaling expression 
\begin{equation}
\chi _{sp}^R({\bf q+Q}_d{\bf ,}\omega )\approx \xi ^{\gamma /\nu }X\left( 
{\bf q}\xi ,\omega \xi ^z\right)  \label{GeneralScalingChi}
\end{equation}
where $X\left( {\bf q}\xi ,\omega \xi ^z\right) $ is a scaling function, we
immediately have the announced results, 
\begin{equation}
\frac \gamma \nu =2\quad ;\quad z=2.
\end{equation}
The Fisher scaling law $\eta =2-\frac \gamma \nu $ shows that the anomalous
exponent $\eta $ vanishes as in mean-field theory. In the following
paragraphs, we compute the remaining exponent $\nu $ to show that above four
dimensions we do recover mean-field theory $\nu =1/2$ while for $2<d<4$, we
have the $n\rightarrow \infty $ result $\nu =1/\left( d-2\right) $.

\subsubsection{Exponent $\nu $ in $2<d<4$ and equivalence to spherical $%
\left( n\rightarrow \infty \right) $ model.}

The correlation length exponent is determined by solving self-consistently
Eqs.(\ref{sumSpin}) and (\ref{Usp}) for the quantity $\langle n_{\uparrow
}n_{\downarrow }\rangle =U_{sp}/U$. In general, we do this numerically using
some technical tricks that are discussed below. With this procedure, no
arbitrariness is left in the cutoffs, that are entirely determined from the
microscopic Hubbard model. However, to study analytically the critical
behavior, we notice that there is a crossover temperature $T_X$ below which
the presence of the small energy scale $\delta U$ $<<T$ makes the zero
Matsubara frequency component in the sum rule Eq.(\ref{sumSpin}) much larger
than all the others. This is the {\it renormalized classical regime}
discussed above. Its existence is a manifestation of critical-slowing down, $%
\omega \sim \xi ^{-2}$ $\sim \delta U$ near a phase transition. Using the
approximate Lorentzian form Eq.(\ref{chiRPA}) for the $i\omega =iq_n=0$
component we rewrite Eq.(\ref{sumSpin}) as follows, after a trivial shift of
integration variables, 
\begin{equation}
\tilde{\sigma}^2=\frac{2T}{U_{sp}\xi _0^2}\int \frac{d^dq}{(2\pi )^d}\frac{%
\xi ^2}{1+q^2\xi ^2}.  \label{Consistency}
\end{equation}
In this equation, 
\begin{equation}
\tilde{\sigma}^2=n-2\langle n_{\uparrow }n_{\downarrow }\rangle -C\leq 1
\label{sigma2-2}
\end{equation}
is the local moment $n-2\langle n_{\uparrow }n_{\downarrow }\rangle $ minus
corrections $C$ that come from the sum over non-zero Matsubara frequencies
(quantum effects) and from the terms neglected in the Lorentzian
approximation, namely those coming from short distances $({\bf q-Q)}^2\gg
\xi ^{-2}$.

Contrary to the strong-coupling case, and contrary to more usual approaches%
\cite{Sachdev}, $\tilde{\sigma}^2$ {\it here is temperature dependent}
because both $\langle n_{\uparrow }n_{\downarrow }\rangle $ and $C$ are.
Nevertheless, to find the critical behavior analytically, it suffices to
notice that this dependence is regular. In fact, we have that when $T\simeq
T_X$, the double occupancy can be approximated by $\left\langle n_{\uparrow
}n_{\downarrow }\right\rangle =U_{sp}/U\approx U_{mf,c}/U$ when $\delta
U\rightarrow 0$.

At the N\'{e}el temperature, $T_{N\text{, }}$ the correlation length
diverges, $\xi =\infty ,$ so that Eq.(\ref{Consistency}) determines the
N\'{e}el temperature through 
\begin{equation}
\tilde{\sigma}^2=\frac{2T_N}{U_{sp}\xi _0^2}\int \frac{d^dq}{(2\pi )^d}\frac 
1{q^2}.  \label{Neel}
\end{equation}
The wave vector integration is cutoff at large $q$ by the Brillouin zone $%
(-\pi <q_i<\pi $ for any component $q_i)$ so that the only divergence occurs
from $q=0$ in $d\leq 2$ ($q=0\rightarrow q={\bf Q}_d$ in the original
integration variables). Since the left-hand side of this Eq.(\ref{Neel}) is
finite, this divergence prevents the existence of a finite-temperature
antiferromagnetic phase transition in two dimensions or less.

To find the correlation length exponent in $2<d<4$, one rewrites Eq.(\ref
{Consistency}) in the form, 
\begin{equation}
\tilde{\sigma}^2=\frac{2T}{U_{sp}\xi _0^2}\int \frac{d^dq}{(2\pi )^d}\left[ 
\frac{\xi ^2}{1+q^2\xi ^2}-\frac 1{q^2}\right] +\frac{2T}{U_{sp}\xi _0^2}%
\int \frac{d^dq}{(2\pi )^d}\left[ \frac 1{q^2}\right] .
\end{equation}
Using the expression for the N\'{e}el temperature Eq.(\ref{Neel}), this last
expression becomes, 
\begin{equation}
\tilde{\sigma}^2\left( 1-\frac T{T_N}\right) =\frac{2T}{U_{sp}\xi _0^2}\xi
^{2-d}\int \frac{d^d\left( q\xi \right) }{(2\pi )^d}\left[ \frac 1{1+q^2\xi
^2}-\frac 1{q^2\xi ^2}\right] .  \label{XsiT-TN}
\end{equation}
Since the integral converges at $q\xi \rightarrow \infty $ for $2<d<4$, it
can be replaced by a $\xi $-independent negative number and one finds, 
\begin{equation}
\xi \sim \left( \frac T{T_N}-1\right) ^{1/\left( 2-d\right) }\sim \left( 
\frac T{T_N}-1\right) ^{-\nu }
\end{equation}
which gives, 
\begin{equation}
\nu =\frac 1{d-2}.
\end{equation}
This exponent and $\gamma /\nu $ $=2$, found in the previous section, are
the same as the one for the Berlin-Kac spherical model\cite{Stanley} or
equivalently for the generalized Heisenberg model where spins are $n$
components vectors and $n\rightarrow \infty $. In three dimensions, this
leads to 
\begin{equation}
\nu =1,\quad \gamma =2,\quad \alpha =-1,\quad \beta ={\frac 12},\quad \eta
=0\ \text{{\rm and}}\ \delta =5.  \label{exponents}
\end{equation}
For comparison, numerical results\cite{Pfeuty} for the $3D$ Heisenberg $%
\left( n=3\right) $ model give $\nu \sim 0.7$ and $\gamma \sim 1.4$.

Above $d=4$, one recovers the mean-field results $\gamma =2\nu $ and $\nu
=1/2$. This last result follows from the fact that in $d>4$, the integral in
Eq.(\ref{XsiT-TN}) is dominated by the large momentum cutoff so that for $%
\xi >>1$, $\left( 1-\frac T{T_N}\right) \sim \xi ^{-2}\int d^dq/q^4$.

\subsubsection{Two-dimensional case}

We have already proven in the last subsection that the transition
temperature vanishes in two dimensions. The correlation length may be found%
\cite{Vilk} in the renormalized classical regime directly by performing the
integral Eq.(\ref{Consistency}) in $d=2$,

\begin{equation}
\xi =\xi _0\left( U_{sp}/\delta U\right) ^{\frac 12}\sim \Lambda ^{-1}\exp
(\pi \tilde{\sigma}^2\xi _0^2U_{sp}/T)  \label{expo}
\end{equation}
where $\Lambda \sim \pi $ is usually of the order of the size of the
Brillouin zone, but not always as we discuss below.

In $d=2$, we call $T_X$ the temperature at which $\delta U\ $is much smaller
than temperature and the magnetic correlation length $\xi $ grows
exponentially. While in higher dimensions a phase transition occurs at
finite temperature, in $d=2$ the critical regime with an exponentially
increasing $\xi $ extends all the way to zero temperature. For example, the
temperature $T_X$ is plotted as a function of filling in the two-dimensional
nearest-neighbor Hubbard model for $U=2.5$ in Fig.1 of Ref.\cite{Vilk}. In
this reference, $T_X$ is called a quasi-critical temperature. We stress that
there is a range of fillings near half-filling where at $T_X$ it is the
antiferromagnetic wave vector that grows, despite the fact that at
zero-temperature the phase transition would be at an incommensurate wave
vector.

The exponential growth of the two-dimensional $\xi $ clearly suggests that
small $3D$ effects existing in real systems may stabilize long-range order
at ${\bf Q}_{d=3}$, before $T=0$. We later characterize the crossover driven
by a small $3D$ hopping parameter $t_{\Vert }\ll t_{\bot }$ from
two-dimensional critical behavior to three-dimensional critical behavior.
But before, we comment on the two-dimensional quantum-critical regime and on
peculiarities induced by nesting in the renormalized-classical regime.

\subsection{Quantum-critical regime}

When there is a critical value of the interaction $U_c$ {\it at zero
temperature} where one finds a paramagnet for $U<U_c$ and an antiferromagnet
for $U>U_c$, then the $T=0$, $U=U_c$ point of the phase diagram is a quantum
critical point.\cite{Hertz} The vicinity of this point in two dimensions has
been studied again recently\cite{Sachdev}. In order to study such a regime
within the Hubbard model at half-filling, one must introduce
next-nearest-neighbor hopping since $U_c\left( T=0\right) =0$ at this
filling in the nearest-neighbor model. One finds that the TPSC approach has
precisely the $n\rightarrow \infty $ model A or model B quantum critical
behavior\cite{Sachdev}, depending on the specific microscopic model. In
particular, $\xi \ $scales as $1/T$ as one approaches the two-dimensional
quantum critical point from finite temperature. Again, in the TPSC approach
the cutoffs are specified without ambiguity. Model C, the perfect nesting
case, is relevant only to the renormalized-classical case, as we now discuss.

\subsection{Peculiarities induced by perfect nesting in the
renormalized-classical regime, especially in $d=2$.}

The dispersion relation of the nearest-neighbor Hubbard model on hypercubic
lattices in arbitrary dimension satisfies $\epsilon _{{\bf k+Q}_d}=-\epsilon
_{{\bf k}}$. Furthermore, at half-filling the particle-hole symmetry implies
that the Fermi surface is fully nested, namely $\mu =0$ so that the equality 
$\epsilon _{{\bf k+Q}_d}-\mu =-\left( \epsilon _{{\bf k}}-\mu \right) $ is
satisfied for all wave vectors ${\bf k}$. Slightly away from half-filling,
nesting in the form $\epsilon _{{\bf k+Q}_d}-\mu \sim -\left( \epsilon _{%
{\bf k}}-\mu \right) $ is also a good {\it approximation} at finite
temperature as long as $T>\mu $, as discussed above. When there is perfect
nesting, the zero-temperature critical interaction vanishes $\left(
U_c=0\right) $. Hence the fully nested Fermi surface, referred to as Model C
above, does not have the simple quantum-critical point described in the
previous sub-section.

When there is perfect nesting, the microscopic interaction-independent
quantities $\xi _0^2$ and $\tau _0$ have a peculiar temperature dependence.
This occurs because they are derivatives of the susceptibility which itself
contains logarithmic singularities in the zero-temperature limit. These
quantities are evaluated in two dimensions and in the quasi two-dimensional
case in Appendix A. Dimensional arguments that follow simply from this
appendix show that in $d>2$ 
\begin{equation}
\xi _0^2\sim 1/\left( T^2\ln T^{-1}\right)  \label{xsi0T}
\end{equation}
\begin{equation}
\tau _0\sim 1/\left( T\ln T^{-1}\right) .
\end{equation}
In $d=2$, the $\ln T^{-1}$ is replaced by $\ln ^2T^{-1}$.\cite{NoteT3}

By contrast, in the case of second-neighbor hopping, nesting is lost and the
above quantities are temperature independent for a wide range of values of
the second-neighbor hopping constant. The above temperature dependencies are
then a special property of nesting. In $d>2$ however, the above temperature
dependencies are completely negligible in the critical regime since near the
phase transition one can replace $T$ in the above expressions by $T_N$.

The only issue then is in two dimensions where the phase transition occurs
at zero temperature. Even neglecting logarithms for the moment, one sees
that since $\xi _0^2$ scales as $1/T^2$ over a wide temperature range the
correlation length in Eq.(\ref{expo}) scales as $\exp \left( cst/T^3\right) $%
. By contrast, in strong coupling, or in the non-nesting case of the
weak-coupling limit, the correlation length scales as $\exp (cst/T)$.

The $\exp \left( cst/\left( T^3\ln ^2T^{-1}\right) \right) $ behavior is
however largely an unsolved problem. Indeed, in the critical regime in two
dimensions, fluctuations remove the quasiparticle peak and replace it by
precursors of the antiferromagnetic bands, as shown in Ref.\cite{Vilk2}. It
is possible then that, in this regime, a more self-consistent treatment
would lead to $\xi _0^2$ independent of temperature, as in the strong
coupling case or the non-nested weak-coupling case. It is also likely that
there will be an intermediate temperature range where the $\exp \left(
cst/\left( T^3\ln ^2T^{-1}\right) \right) $ regime prevails, even if deep in
the critical regime self-consistency leads to $\exp \left( cst/T\right) $
behavior.

It is important to recall that in practical calculations in the TPSC
approach, one obtains a numerical value for the correlation length without
adjustable parameter. For example in Fig.1 we present the temperature
dependence of the correlation length for the two-dimensional
nearest-neighbor Hubbard model. As discussed in Appendix A, in this case 
\begin{equation}
\xi _0^2\simeq {0.021U}_{mf,c}t_{\bot }^2a_{\bot }^2/T^2  \label{NumXsiZero}
\end{equation}
and $U_{sp}\simeq {U}_{mf,c}$ so that from the slope of the plot and from
Eq.(\ref{expo}) one finds $\tilde{\sigma}^2\simeq 0.21$. From the plot we
can also extract $\Lambda ^{-1}\simeq 0.022$ so that $\xi $ is known without
adjustable parameter. Appendix B explains physically the orders of magnitude
taken by $\tilde{\sigma}^2$ and $\Lambda ^{-1}$ in this model. Similar
calculations can be done for arbitrary band structure. In strong-coupling
calculations,\cite{Chakravarty}\cite{Chubukov} one obtains $\xi \sim \Lambda
^{-1}\exp (2\pi \rho _S/T)$ with $\rho _S$ a cutoff-dependent quantity that
can be evaluated only with Monte Carlo simulations.

Another consequence of the temperature behavior of $\xi _0$ in Eq.(\ref
{xsi0T}) is that {\it above} $T_X$ there is a range of temperatures for
which the antiferromagnetic correlation length scales as $\xi $ $\sim \xi
_0\sim 1/T$. This behavior should not be confused with quantum-critical
behavior, even though the power-law scaling of the correlation length is the
same. Indeed, one finds that the argument of the exponential in Eq.(\ref
{expo}) is larger than unity in the corresponding regime while in the
quantum-critical regime the argument of the exponential should be much less
than unity.\cite{Sachdev} In fact the temperature dependence of the
staggered susceptibility for $T>T_X$ is also-different from the quantum
critical result.

\section{Quasi two-dimensional systems: Renormalized classical crossover
from $d=2$ to $d=3$.}

The general discussion of universality in the renormalized-classical
crossover from $d=2$ to $d=3$ appears in Appendices C and D. In the present
section, we first clarify the various regimes of crossover, according to
whether or not single-particle coherence in the third dimension is
established before the phase transition. Then, we go on to discuss the case $%
t_{\Vert }\ll T_N<T_X$ where the SDW instability occurs before interplane
single-particle coherence is established. More specifically, we find the
scaling of the N\'{e}el temperature with $t_{\bot }/t_{\Vert }$ as well as
the size of the three-dimensional critical regime with the corresponding
exponents, showing that the results are those of the $n\rightarrow \infty $
limit. We restore the lattice spacing units $a_{\Vert }$ along the
three-dimensional axis and $a_{\bot }$ in the planes. We assume however that
the ratio $a_{\Vert }/a_{\bot }$ is usually of order unity and numerical
calculations are done for $a_{\Vert }/a_{\bot }=1$

\subsection{One-particle and two-particle crossover from $d=2$ to $d=3.$}

We consider in this section the highly anisotropic situation where hopping
between planes $t_{\Vert }$ is much smaller than in-plane hopping, $t_{\bot
} $, 
\begin{equation}
t_{\Vert }\ll t_{\bot }
\end{equation}
as might occur in the high-temperature superconductor parent compound $%
La_2CuO_4$. In this case, we have that the three-dimensional transition
temperature to long-range order $T_N$ is always less than the crossover
temperature $T_X$ to the characteristic exponential behavior of the
correlation length in two dimensions 
\begin{equation}
T_N<T_X.
\end{equation}
This is so because the microscopic in-plane $\xi _0^{\perp }$ and out of
plane $\xi _0^{\Vert }$ lengths satisfy $\xi _0^{\perp }$ $\gg \xi _0^{\Vert
}$.

The crossover temperature to two-dimensional behavior for itinerant
antiferromagnet always satisfies 
\begin{equation}
T_X<t_{\bot }.
\end{equation}
Two limiting cases are then possible, depending on interaction and on
hopping parallel to the three-dimensional axis $t_{\parallel }$:

{\em a) Weak coupling or small anisotropy limit:} 
\begin{equation}
T_N<T_X\ll t_{\Vert }.
\end{equation}
In this case, when the transition to three-dimensional behavior occurs the
three-dimensional Fermi surface is relevant since the thermal de Broglie
wavelength $v_F^{\Vert }/\left( \pi T\right) \sim $ $t_{\Vert }a_{\Vert }/T$
is larger than the distance between planes. In other words, the
three-dimensional band structure is relevant to the behavior of {\it %
single-particle} propagators in Matsubara frequencies before the phase
transition occurs. Fermions are quantum-mechanically coherent over more than
a single plane and nesting generally plays a role in the value of the
ordering wave vector. The crossover from two to three dimensional {\it %
critical} behavior would occur in a manner analogous to the anisotropic
Heisenberg model\cite{Soukoulis}\cite{Singh}\cite{Tesanovic}\cite{Kosterlitz}%
.

{\em b) Intermediate coupling or very large anisotropy:} 
\begin{equation}
t_{\Vert }\ll T_N<T_X.  \label{IncoherentRegime}
\end{equation}
Here, long-range order is established before the single-particle coherence
occurs between planes. A phase transition occurs only because of {\it %
two-particle} (or particle-hole) coherent hopping. When the phase transition
occurs, thermal fluctuations are still large enough that coherent
single-particle band motion in the parallel direction has not occured yet.
There are several ways of explaining physically what this last statement
means. For example, it is clear that when $t_{\Vert }\ll T,$ features of the
band structure in the parallel direction are irrelevant for single-particle
properties since the first Matsubara frequency is larger than the bandwidth
in that direction. The motion {\it between} planes is still in that sense
quasi-classical when the phase transition occurs. Another way of saying what
this means is that the thermally induced uncertainty in the parallel wave
vector is equal to the extent of the Brillouin zone in that direction,
corresponding, via the uncertainty principle, to a confinement within each
plane.

We do not discuss the intermediate case $T_N<t_{\Vert }<T_X$ but concentrate
instead on the very large anisotropy - intermediate coupling limit just
introduced.

\subsection{Numerical solutions}

The scaling behavior of the N\'{e}el temperature $T_N$ and of the
three-dimensional crossover temperature are derived in the following two
sections. We first present the numerical results obtained from the solution
of the self-consistency relations Eqs.(\ref{sumSpin}) and (\ref{Usp}). The
numerical integration in Eq.(\ref{sumSpin}) is made possible by rewriting
this equation in the form,

\begin{equation}
n-2\langle n_{\uparrow }n_{\downarrow }\rangle =Ta_{\Vert }a_{\bot }^2\int 
\frac{d^3q}{\left( 2\pi \right) ^3}\sum_{iq_n}\left[ \chi _{sp}\left( {\bf q,%
}iq_n\right) -\chi _{sp}^{as}\left( {\bf q,}0\right) \delta _{n,0}\right]
+Ta_{\Vert }a_{\bot }^2\int \frac{d^3q}{\left( 2\pi \right) ^3}\chi
_{sp}^{as}\left( {\bf q,}0\right) .  \label{SelfQ2d}
\end{equation}
The sum over large Matsubara frequencies can be approximated by an integral
in a controlled manner. The subtraction in the first integral removes
singularities of the integrand and makes the integral well behaved. Since
the transition occurs in the single-particle incoherent regime Eq.(\ref
{IncoherentRegime}), the integrand in square brackets is independent of $%
q_{\Vert }$ above the N\'{e}el temperature. All quantities involving $%
t_{\parallel }$ come from the second integral over the asymptotic expression
for the susceptibility 
\begin{equation}
\chi _{sp}^{as}\left( {\bf q+Q}_3{\bf ,}0\right) \equiv \frac 2{\delta U}%
\frac 1{1+\xi _{\parallel }^2q_{\parallel }^2+\xi _{\perp }^2q_{\perp }^2}
\label{ChiAs}
\end{equation}
where 
\begin{equation}
\xi _{\perp }\equiv \xi _0^{\perp }(U_{sp}/\delta U)^{1/2},  \label{ksi2per}
\end{equation}
\begin{equation}
\xi _{\parallel }\equiv \xi _0^{\parallel }(U_{sp}/\delta U)^{1/2}.
\label{ksi2par}
\end{equation}
To have sufficient precision for large two-dimensional correlation lengths,
it is important to evaluate analytically $\xi _0^{\perp }$, $\xi
_0^{\parallel }$, as well as the integral of Eq.(\ref{ChiAs}) appearing in
the consistency equation Eq.(\ref{SelfQ2d}). This is done respectively in
Appendices A and C. To perform the second derivatives in the definition of $%
\xi _0^{\perp }$, $\xi _0^{\parallel }$, we expand $\chi _0({\bf Q}_3)$ in
powers of $t_{\parallel }/T$, keeping only the first non-zero term: thus $%
\xi _0^{\perp }$ does not differ from the one already presented in Eq. (\ref
{NumXsiZero}). It is shown in Appendix A that over a wide range of
temperatures we have 
\begin{equation}
\frac{\partial ^2\chi _0\left( {\bf Q}_3\right) }{\partial q_{\bot }^2}\sim
a_{\bot }^2\left[ \frac{t_{\bot }^2}{T^2}+O(t_{\Vert }/T)\right] ,
\label{derivx}
\end{equation}
and 
\begin{equation}
\frac{\partial ^2\chi _0\left( {\bf Q}_3\right) }{\partial q_{\Vert }^2}\sim
a_{\Vert }^2\left[ \frac{t_{\Vert }^2}{T^2}+O(t_{\Vert }^3/T^3)\right] .
\label{derivz}
\end{equation}
The inter-plane hopping $t_{\parallel }$ in Eqs.(\ref{SelfQ2d}) and (\ref
{ChiAs}) occurs explicitly only in $\xi _0^{\parallel }$ and the above
results imply that, 
\begin{equation}
\frac{\xi _0^{\parallel }}{\xi _0^{\perp }}\sim \frac{\xi _{\parallel }}{\xi
_{\perp }}\sim \frac{t_{\parallel }a_{\Vert }}{t_{\bot }a_{\bot }}.
\label{tzot}
\end{equation}

We present numerical results for the nearest-neighbor Hubbard model in units
where $a_{\Vert }=a_{\bot }=1$ and $t_{\bot }=1$. The value of $T_N\left(
t_{\parallel }\right) $ appears in Figs.2 and 3 for $U=4$. In Fig.2, we
clearly see that $T_N$ becomes almost equal to $T_X\approx 0.2$ for $%
t_{\Vert }$ still quite small. The scaling of $T_N\left( t_{\parallel
}\right) $ shown in Fig.3 is explained in the following subsection. Fig.4
shows the variation of the in-plane correlation length $\xi _{\perp }$ as a
function of temperature for various $t_{\parallel }$, again for $U=4$. For
the purely two-dimensional case $t_{\parallel }$ $=0$, one can observe for $%
T\leq T_X\simeq 0.2$ the exponential behavior mentioned in the preceding
section. At a temperature about $0.16$, the in-plane correlation length $\xi
_{\bot }$ is already as much as $10^3$ (in units where lattice space is
unity). For $10^{-3}\leq t_{\Vert }\leq 10^{-1}$, $\xi _{\bot }$ diverges at
the N\'{e}el temperature located in the narrow range $0.16\leq T_N\leq 0.2$.
For $t_{\parallel }$ $\neq 0$, it is also clear that the crossover to
three-dimensional behavior occurs in an extremely narrow temperature range.
This is explained below. Note that the last few curves on the right-hand
side are at the limit of validity of our approximations.

We note that for $U=4t$, we find that at $T=T_N$ the local moment is equal
to three quarters of the full moment in the atomic limit, {\it i.e.} $%
n-2\langle n_{\uparrow }\rangle \langle n_{\downarrow }\rangle g_{\uparrow
\downarrow }(0)=0.75$,{\it \ } $g_{\uparrow \downarrow }(0)=\langle
n_{\uparrow }n_{\downarrow }\rangle /\langle n_{\uparrow }\rangle \langle
n_{\downarrow }\rangle .$ This number is only weakly dependent on
temperature in the range studied.

\subsection{Dependence of the N\'{e}el temperature $T_N\,$on $t_{\Vert }$.
Crossover exponent.}

From the discussion of the previous section, Eqs.(\ref{SelfQ2d})(\ref{ChiAs}%
)(\ref{Consistency}) we know that the singular part of the self-consistency
condition may be written in the following form in the quasi-two dimensional
case,

\begin{equation}
\tilde{\sigma}^2=\frac{2Ta_{\Vert }a_{\bot }^2}{U_{sp}\left( \xi _0^{\perp
}\right) ^2}\int \frac{d^3q}{(2\pi )^3}\frac 1{q_{\perp }^2+\xi _{\perp
}^{-2}+\left( \frac{\xi ^{\parallel }}{\xi ^{\perp }}\right) ^2q_{\parallel
}^2}.  \label{sigma2-3}
\end{equation}
The integral can be done exactly, as in Appendix C, and all the results
obtained in this subsection and the following one can be obtained from
limiting cases of this general analytical result, as shown in Appendix D.
Here we make approximations directly on the integrals since this makes the
Physics of the results more transparent. Although arbitrary cutoffs appear
in the analytical expressions, we re-emphasize that in the numerical
calculations of the previous section the cutoffs are simply given by the
Brillouin zone and there is no arbitrary scale in the results.

At $T_N$, we have $\xi _{\perp }^{-2}=0$. Furthermore, from Eqs.(\ref
{ksi2per}) and (\ref{ksi2par}) we find $\xi _{\Vert }/\xi _{\perp }=\xi
_0^{\parallel }/\xi _0^{\perp }$ so that the above integral Eq.(\ref
{sigma2-3}) takes the form 
\begin{equation}
\tilde{\sigma}^2=\frac{2T_Na_{\Vert }a_{\bot }^2}{U_{sp}\left( \xi _0^{\perp
}\right) ^2}\int \frac{d^2q_{\bot }}{(2\pi )^2}\left[ \int \frac{dq_{\Vert }%
}{2\pi }\frac 1{q_{\perp }^2+\left( \frac{\xi _0^{\parallel }}{\xi _0^{\perp
}}\right) ^2q_{\parallel }^2}\right] .
\end{equation}
Using the mean-value theorem for the integral over $q_{\Vert }$ we have 
\begin{equation}
\tilde{\sigma}^2=\frac{2T_Na_{\bot }^2}{U_{sp}\left( \xi _0^{\perp }\right)
^2}\int \frac{d^2q_{\bot }}{(2\pi )^2}\left[ \frac 1{q_{\perp }^2+\left( 
\frac{\xi _0^{\parallel }}{\xi _0^{\perp }}\right) ^2\widetilde{\Lambda }^2}%
\right]
\end{equation}
where $\widetilde{\Lambda }$ is a constant that we need not specify. It is
contained in the range $0<$ $\left| \widetilde{\Lambda }\right| <\pi
/a_{\Vert }$. The above integral is the same as the one that determines the
correlation length in two dimensions Eq.(\ref{Consistency}), hence at $T_N$
we have that 
\begin{equation}
\xi _{\perp }^{-2}\left( T_N\right) =\left( \frac{\xi _0^{\parallel }}{\xi
_0^{\perp }}\right) ^2\widetilde{\Lambda }^2\propto \left( \frac{t_{\Vert
}a_{\Vert }}{t_{\bot }a_{\bot }}\right) ^2\widetilde{\Lambda }^2.
\label{XsiTn}
\end{equation}
Comparing with the general theory of Appendix D where it is argued that $\xi
_{\bot }^{-\phi /\nu }\left( T_N\right) \sim t_{\Vert }^2$, we see that $%
\phi /\nu =2$. In other words, $\phi /\nu =\gamma /\nu =2$ and the crossover
exponent\cite{Pfeuty} $\phi $ is here equal to$\ \gamma $ as is usually the
case in the $n\rightarrow \infty $ model.\cite{Riedel} We obtain, using the
expression (\ref{expo}) for the correlation length in two dimensions, 
\begin{equation}
\frac 1{T_N}=\frac{a_{\bot }^2}{\pi \widetilde{\sigma }^2U_{sp}\left( \xi
_0^{\perp }\right) ^2}\left[ \ln \frac{t_{\bot }}{t_{\Vert }}+c\right]
\label{1otn}
\end{equation}
where $c$ is a non-universal constant of order unity.

In the special case of perfect nesting (half-filled nearest-neighbor hopping
model), the microscopic length $\xi _0^{\perp }$ is temperature dependent,
as shown in Appendix A 
\begin{equation}
\left( \xi _0^{\perp }\right) ^2\sim a_{\bot }^2\frac{0.085}{T^2}\frac 1{%
2\chi \left( {\bf Q}_2\right) }.
\end{equation}
Using this result as well as $U_{mf,c}\equiv 2/\chi \left( {\bf Q}_2\right)
\approx U_{sp}$ at $T_X\approx T_N$ gives the scaling illustrated in Fig.3
for the case $a_{\bot }/a_{\Vert }=1$ namely 
\begin{equation}
\frac 1{T_N}\sim \frac{T_N^2}{U_{mf,c}^2}\left| \ln \frac{t_{\Vert }}{%
t_{\bot }}\right|
\end{equation}

The logarithmic behavior in Eq.(\ref{1otn}) is typical of systems that
undergo a dimensional crossover from their lower critical dimension. For
example, the analog of Eq.(\ref{XsiTn}) in the anisotropic Heisenberg case
would read,\cite{Soukoulis} 
\begin{equation}
\xi _{\perp }^{-2}\left( T_N\right) \sim \left( \frac{J_{\Vert }}{J_{\bot }}%
\right) \widetilde{\Lambda }^2  \label{AH}
\end{equation}
leading to\cite{Soukoulis} $T_N^{-1}\sim \ln \left( \frac{J_{\Vert }}{%
J_{\bot }}\right) .$ The above results Eqs.(\ref{XsiTn}) and (\ref{AH}) are
suggested by the simple RPA-like form $\chi _{3d}\sim \chi _{2d}/\left(
1-J_{\Vert }\chi _{2d}\right) $ with $J_{\Vert }\chi _{2d}\sim 1$ at the
transition and $\chi _{2d}\sim \xi _{\perp }^{\gamma /\nu }\sim \xi _{\perp
}^2\sim \exp \left( J_{\bot }cst/T\right) $. As in the previous section, the
quantity $U_{sp}\left( \xi _0^{\perp ,\Vert }\right) ^2$ plays a role
analogous to the exchange constants $J_{\bot \parallel ,}$. In the perfect
nesting case, these effective exchange constants would be temperature
dependent since $U_{sp}\left( \xi _0^{\perp ,\Vert }\right) ^2$ $\sim
U_{sp}\left( t_{\perp ,\Vert }\right) ^2/T^2$. Note that in the crossover
from one-dimensional Luttinger liquid behavior to three-dimensional
long-range order, the effective exchange constant $J_{\bot }$ also scales%
\cite{BourbonnaisCaron}\cite{Boies} as $ut_{\bot }^2/T^2$, with $u$ a
running coupling constant. The one-dimensional Fermi surface is always
nested.

\subsection{Size of the three-dimensional critical region.}

The singular temperature dependence of the correlation length is obtained
from the equation

\begin{equation}
\tilde{\sigma}^2=\frac{2Ta_{\Vert }a_{\bot }^2}{\delta U}\int \frac{d^3q}{%
(2\pi )^3}\frac 1{1+\xi _{\perp }^2q_{\perp }^2+\xi _{\parallel
}^2q_{\parallel }^2}.  \label{sigma2-4}
\end{equation}
Since the ratio $\xi _{\Vert }/\xi _{\perp }=\xi _0^{\parallel }/\xi
_0^{\perp }$ is temperature independent, a simple change of integration
variables shows that near $T_N$ the scaling of both correlation lengths with
temperature is identical to the isotropic three-dimensional case. In other
words, the critical behavior near the phase transition is that of the
three-dimensional system. However, as one increases the temperature away
from $T_N$, the correlation lengths can decrease until $\xi _{\parallel }\ll
a_{\Vert }$ while at the same time $\xi _{\perp }\gg a_{\bot }$. When $\xi
_{\parallel }\ll a_{\Vert }$, the integral Eq.(\ref{sigma2-4}) is
essentially two-dimensional and for $\xi _{\perp }\gg a_{\bot }$ one should
observe the characteristic exponential temperature dependence of the
two-dimensional correlation length.

As usual the definition of crossover contains some arbitrariness, so let us
choose 
\begin{equation}
\xi _{\parallel }\left( T^{*}\right) =a_{\Vert }  \label{Xsi=1}
\end{equation}
as the definition of the crossover temperature $T^{*}$ between $d=2$ and $%
d=3 $ critical behavior. In that regime, the correlation length Eq.(\ref
{Xsi=1}) scales with temperature as in the $d=2$ regime Eq.(\ref{expo})
except that, as argued before, $\xi _{\Vert }$ is smaller by a factor $%
\left( \xi _0^{\Vert }/\xi _0^{\bot }\right) =\left( t_{\Vert }a_{\Vert
}\right) /\left( t_{\bot }a_{\bot }\right) $, hence we obtain for $T^{*}$ 
\begin{equation}
\frac 1{T^{*}}=\frac{a_{\bot }^2}{\pi \widetilde{\sigma }^2U_{sp}\left( \xi
_0^{\bot }\right) ^2}\left[ \ln \left( \frac{t_{\bot }}{t_{\Vert }}\right)
+c^{\prime }\right]  \label{CrossExp}
\end{equation}
where $c^{\prime }$ is a non-universal constant of order unity. The size of
the crossover region is thus 
\begin{equation}
1-\frac{T_N}{T^{*}}=T_N\frac{a_{\bot }^2}{\pi \widetilde{\sigma }%
^2U_{sp}\left( \xi _0^{\bot }\right) ^2}\left( c-c^{\prime }\right) =\frac{%
\left( c-c^{\prime }\right) }{\ln \left( \frac{t_{\bot }}{t_{\Vert }}\right)
+c}\sim \frac 1{\ln \left( \frac{t_{\bot }}{t_{\Vert }}\right) }.
\label{SizeCrossover}
\end{equation}
The above results Eqs.(\ref{Xsi=1}) to (\ref{SizeCrossover}) are as expected
from the usual theory of critical phenomena exposed in Appendix D. In
particular, the scaling of $T^{*}$ with $t_{\Vert }/t_{\bot }$ is the same
as that of $T_N.$ The smallness of the crossover region from $d=2$ to $d=3$
critical behavior in Fig.4 follows from the above considerations. The
smaller is $T_N$, the smaller is $T^{*}$. The above situation should be
contrasted with the problem of crossover from $d=3$ to $d=2$ in Helium
films, studied by Fisher and Barber\cite{Fisher} many years ago. In that
case, power law scaling occurred everywhere, giving quite different
expressions for the scaling of $T^{*}$ and $T_N$.

Given $\xi _{\parallel }\left( T^{*}\right) =\xi _0^{\parallel }\left(
T^{*}\right) \left( U_{sp}/\delta U\left( T^{*}\right) \right) ^{1/2}$, and $%
\xi _0^{\parallel }\sim t_{\parallel }$, the above relation $\xi _{\parallel
}\left( T^{*}\right) =a_{\Vert }$ means that $\delta U\left( T^{*}\right) $
should scale as $t_{\parallel }^2$. Similarly we should have $\delta U\left(
T_N\right) \sim t_{\parallel }^2.$ We checked numerically\cite{DareThese}
that the scaling with $t_{\Vert }$ holds for $t_{\Vert }<0.05$ in the
half-filled nearest-neighbor model with $U=4t_{\perp }.$

\section{Conclusion}

We have shown that the TPSC approach allows one to study all aspects of
nearly antiferromagnetic itinerant electrons in one-band Hubbard models. The
method is in quantitative agreement with Monte Carlo simulations in the
non-critical regime\cite{Vilk}\cite{Veilleux} while in the critical regime,
(renormalized classical or quantum critical) the relatively weak temperature
dependence of the local moment leads to the same critical behavior as
strong-coupling models to leading order in the $1/n$ expansion, namely in
the $n\rightarrow \infty $ limit. There is no arbitrary cutoff so that all
results can be obtained as a function of lattice spacing, hopping integral
and interaction parameter. Fermi surface effects are apparent, in particular
in the case of perfect nesting where the two-dimensional
renormalized-classical correlation length diverges as $\exp \left(
cst/\left( T^3\ln ^2T\right) \right) $ instead of $\exp \left( cst/T\right) $%
.

We have applied the method to a detailed study of the renormalized-classical
crossover from two to three dimensions where we have highlighted the
existence of a regime where the three-dimensional N\'{e}el instability
occurs before thermal fluctuations become small enough to allow coherent
single-particle band motion between planes. An analogous phenomenon occurs
in quasi-one dimensional systems\cite{BourbonnaisCaron}\cite{Boies}. The
TPSC approach can be applied to study realistic cases. For $La_2CuO_4$ we
will show in a subsequent publication that with second-neighbor hopping one
can fit experiments on the magnetic structure factor.

The generalization of the TPSC approach beyond leading order in $1/n$ is
left open. Also, the effect of self-energy feedback\cite{Vilk2} on $\exp
\left( cst/\left( T^3\ln ^2T\right) \right) $ behavior of the correlation
length in the two-dimensional nesting case should be cleared in further
studies. Finally, the universal $d=2$ to $d=3$ crossover discussed in
Appendix D should be investigated beyond leading order in $1/n$.

We are indebted to C. Bourbonnais for numerous discussions and key ideas. We
also thank David S\'{e}n\'{e}chal for discussions and D.S. Fisher for
pointing out Ref.\cite{Kosterlitz}. We acknowledge the support of the
Natural Sciences and Engineering Research Council of Canada (NSERC), the
Fonds pour la formation de chercheurs et l'aide \`{a} la recherche from the
Government of Qu\'{e}bec (FCAR) and (A.-M.S.T.) the Canadian Institute for
Advanced Research (CIAR) and the Killam foundation.

\appendix 

\section{$\xi _0^{\Vert ,\bot }$ and $\tau _0$ in the case of nesting}

In this appendix we derive expressions for the out of plane $\xi _0^{\Vert }$
and in-plane $\xi _0^{\bot }$ microscopic lengths, 
\begin{equation}
\xi _0^{\Vert ,\bot }=\frac{-1}{2\chi _0\left( {\bf Q}_d\right) }\left. 
\frac{\partial ^2\chi _0\left( {\bf q,}0\right) }{\partial q_{\Vert ,\bot }^2%
}\right| _{{\bf q=Q}_d}
\end{equation}
as well as for the microscopic relaxation time $\tau _0$ in Eq.(\ref{Gamma0}%
) 
\begin{equation}
\tau _0=\frac 1{\chi _0\left( {\bf Q}_d\right) }\left. \frac{\partial \chi
_0^R\left( {\bf Q}_d{\bf ,}\omega \right) }{\partial i\omega }\right|
_{\omega =0}
\end{equation}
for the quasi two-dimensional antiferromagnet ${\bf Q}_3=(\pi ,\pi ,\pi )$ ,%
{\bf \ }in the regime $t_{\Vert }\ll T_N<T_X$ of Eq.(\ref{IncoherentRegime}%
). We also assume that $\mu \ll T$ so that the maximum of the static
susceptibility is at ${\bf Q}_3$ even away from half-filling.

We start from the retarded Lindhard function in $d$ dimensions 
\begin{equation}  \label{a3}
\chi _0^R({\bf q},\omega )=2a_{\Vert }a_{\bot }^2\int_{BZ}{\frac{d^dk}{(2\pi
)^d}}\ {\frac{f(\epsilon _{{\bf k+q}}-\mu )-f(\epsilon _{{\bf k}}-\mu )}{%
\omega +i\eta -\epsilon _{{\bf k+q}}+\epsilon _{{\bf k}}},}
\end{equation}
where $f$ is the Fermi function, $\mu $ the chemical potential ($\mu =0$ at
half-filling for our Hamiltonian but the expressions quoted here are more
general than for the half-filled case). For nearest-neighbor hopping, we
have the nesting property 
\begin{equation}
\epsilon _{{\bf k+Q}_d}=-\epsilon _{{\bf k}}
\end{equation}
that can be used to rewrite 
\begin{equation}
\chi _0^R({\bf Q}_d,\omega )=2\int {dEN}_d(E)\ {\frac{1-f(E+\mu )-f(E-\mu )}{%
\omega +i\eta +2E}}
\end{equation}
where ${N}_d(E)$ is the single-spin density of states for the given
dimension.

In the limit $\mu \ll T$ we have for the static susceptibility 
\begin{equation}
\chi _0({\bf Q}_d)\equiv \chi _0^R({\bf Q}_3,0)=2\int {dEN}_d(E)\ {\frac{%
1-2f(E)}{2E}}
\end{equation}
so that in two dimensions $\chi _0({\bf Q}_2)\sim \ln ^2\left( t/T\right) $
while in three dimensions $\chi _0({\bf Q}_3)\sim \ln \left( t/T\right) $.
In the quasi two-dimensional case with $t_{\Vert }\ll T$ the two-dimensional
value of $\chi _0$ is an accurate approximation. The numerical values of $%
\chi _0({\bf Q}_d)$ are in practice easy to obtain from numerical
integrations.

For the microscopic relaxation time when $\mu \ll T$ we start from 
\begin{equation}
\mathop{\rm Im}
\chi _0^R({\bf Q}_d,\omega )=\pi {N}_d(\frac \omega 2)\tanh \left( \frac 
\omega {4T}\right) .
\end{equation}
In two dimensions, the logarithmic divergence of the density of states ${N}%
_d(\frac \omega 2)$ at the van Hove singularity makes the zero-frequency
limit of the microscopic relaxation time ill-defined. Nevertheless, van Hove
singularities are usually washed out by lifetime effects in more
self-consistent treatments so that one expects that for $\omega <<T$ one has 
$\left. \partial \chi _0^R\left( {\bf Q}_d{\bf ,}\omega \right) /d\omega
\right| _{\omega =0}\sim 1/T$ leading to the temperature scaling of $\tau
_0\sim \left( T\ln T^{-1}\right) ^{-1}$ in $d>2$ described in the text.

We move on to evaluate analytically the wave vector derivatives in the
regime $t_{\Vert }\ll T_N<T_X$. Keeping for a while a general notation where 
$i$ is some direction $(x,y,$ or $z)$, one can write 
\begin{equation}
{\frac{\partial ^2\chi _0}{\partial q_i^2}}=-8t_i^2a_i^2\int_{BZ}{\frac{d^3k%
}{(2\pi )^3}}{\frac{\partial ^2C}{\partial \epsilon _{{\bf k+q}}^2}}\ \sin
^2(k_i+q_i)-4t_ia_i\int_{BZ}{\frac{d^3k}{(2\pi )^3}}{\frac{\partial C}{%
\partial \epsilon _{{\bf k+q}}}}\cos (k_i+q_i),  \label{a4}
\end{equation}
where 
\[
C(\epsilon _{{\bf k+q}},\epsilon _{{\bf k}})={\frac{f(\epsilon _{{\bf k+q}%
}-\mu )-f(\epsilon _{{\bf k}}-\mu )}{\epsilon _{{\bf k+q}}-\epsilon _{{\bf k}%
}}.} 
\]
Assuming $t_{\Vert }/T\ll 1$, we evaluate second derivatives to the lowest
non-zero term in powers of $t_{\Vert }/T$. For $q_i=q_{\parallel }$, the
leading term in Eq. (\ref{a4}) gives a $t_{\Vert }^2$ contribution if we
keep $t_{\Vert }=0$ in the integrand. The second term gives also to leading
order a quadratic contribution in $t_{\Vert }$. The spread of the Fermi
factors over an energy interval of order $T$ allows us to neglect all other
dependencies in $t_{\Vert }$ and to perform the integral in the third
direction trivially, enabling us to rewrite the remaining integral in terms
of the two-dimensional single-spin density of states $N_2(E)$. After some
algebra we get 
\begin{equation}
{\frac{\partial ^2\chi _0({\bf Q}_{3D})}{\partial q_{\parallel }^2}}%
=-2t_{\Vert }^2a_{\Vert }^2\int_0^{4t}dEN_{2D}(E)\{{\frac{f^{\prime }(E+\mu
)+f^{\prime }(E-\mu )}{E^2}}+{\frac{1-f(E+\mu )-f(E-\mu )}{E^3}}\}+O(({\frac{%
t_{\Vert }}T})^3)  \label{a6}
\end{equation}
where $f^{\prime }$ is the derivative of the Fermi function. Using the
expansion of Fermi functions and derivatives near $E=0$, it can be shown
that the integrand in the preceding equation is finite at finite
temperature. Indeed as $E/T\rightarrow 0$, it behaves as $N_{2D}(E)f^{\prime
\prime \prime }(\mu )$, where $f^{\prime \prime \prime }$ is the third
derivative of the Fermi function. At low temperature, approximating the
integrand by $N_{2D}(E)f^{\prime \prime \prime }(\mu $) over an energy
interval $T$ shows immediately that 
\begin{equation}
{\frac{\partial ^2\chi _0({\bf Q}_{3D})}{\partial q_{\parallel }^2}}\sim
a_{\Vert }^2{\frac{t_{\Vert }^2}{T^2}}.  \label{a7}
\end{equation}
More precisely, $1/T^2$ should be multiplied by a logarithmic correction
that comes from the $2D$ density of states. Numerical integration of (\ref
{a6}) shows that this $t_{\Vert }^2{\frac 1{T^2}}$ behavior occurs on a
wider range of temperature than first expected: $T=0.2t$ is already in this
regime.

We can evaluate the in-plane $q_i=q_{\bot }$ derivative in the same spirit.
This time take $t_{\Vert }=0$ from the start since the leading order is in $%
t_{\bot }^2/T^2\gg $ $t_{\Vert }^2/T^2$. We thus have $\partial ^2\chi _0(%
{\bf Q}_3)/\partial q_{\perp }^2${\ }${\approx \partial ^2\chi _0({\bf Q}%
_2)/\partial q_{\perp }^2}$. After tedious algebra we finally get 
\[
{\frac{\partial ^2\chi _0({\bf Q}_3)}{\partial q_{\perp }^2}}\simeq t_{\bot
}^2a_{\bot }^2\int_0^{4t}dEN_{2D}(E)\left\{ {\frac 12}\left[ f^{\prime
}(E+\mu )+f^{\prime }(E-\mu )\right] +{\frac{1-f(E+\mu )-f(E-\mu )}{2E}}%
\right\} 
\]
\[
-t_{\bot }^2a_{\bot }^2\int_0^{4t}dEM(E)\left\{ {\frac 1E}\left[ f^{\prime
\prime }(E+\mu )+f^{\prime \prime }(E-\mu )\right] \right. 
\]
\begin{equation}
-\left. {\frac{f^{\prime }(E+\mu )+f^{\prime }(E-\mu )}{E^2}}-{\frac{%
1-f(E+\mu )-f(E-\mu )}{E^3}}\right\} ,  \label{a8}
\end{equation}
where the integral 
\begin{equation}
M(E)\equiv \int_{-\pi }^\pi dq_x4\sin ^2q_x\int \frac{d^2k}{\left( 2\pi
\right) ^2}\delta \left( E-\epsilon _{{\bf k}}\right) \delta \left(
q_x-k_x\right)
\end{equation}
can be interpreted as an average over the surface of constant energy $E$ of
the square of the Fermi velocity in the $x$ direction times the density of
states at this energy. It can be evaluated analytically as 
\begin{equation}
M(E)={\frac 2{\pi ^2}}\{2{\bf E}(k^2)-{\frac E{2t_{\bot }}}(1+{\frac E{%
2t_{\bot }}}){\bf F}(k^2)+{\frac{E^2}{4t_{\bot }^2}}{\bf \Pi }(\alpha
^2,k^2)\}.
\end{equation}
Here ${\bf F}(k^2)$, ${\bf E}(k^2)$ and ${\bf \Pi }(\alpha ^2,k^2)$ are
complete elliptic integrals of respectively first, second and third kinds,
with $k^2=1-E^2/\left( 16t_{\bot }^2\right) ,\ \ {\rm and}\ \ \alpha
^2=1-E/\left( 4t_{\bot }\right) $. Again at $E/T=0$ the integrand in Eq.(\ref
{a8}) is well defined, and using Fermi function expansion it can be shown
that at low temperature $\partial ^2\chi _0(Q)/\partial q_{\perp }^2$ scales
as $a_{\bot }^2t_{\bot }^2/T^2$, with this time {\it no} logarithmic
prefactor as before. More precisely we found numerically for a wide range of
temperature (wider than the range studied in the main text) the following
behavior 
\begin{equation}
{\frac{\partial ^2\chi _0({\bf Q}_e)}{\partial q_{\perp }^2}}\simeq {-0.085}%
a_{\bot }^2\frac{t_{\bot }^2}{T^2}  \label{a9}
\end{equation}
and correspondingly, 
\[
\left( \xi _0^{\bot }\right) ^2\equiv \frac{-1}{2\chi _0\left( {\bf Q}%
_d\right) }\left. \frac{\partial ^2\chi _0\left( {\bf q,}0\right) }{\partial
q_{\bot }^2}\right| _{{\bf q=Q}_d}\simeq {0.085}a_{\bot }^2\frac{{U}_{mf,c}}4%
\frac{t_{\bot }^2}{T^2}. 
\]
From Eqs.(\ref{a7}) and (\ref{a9}) and a numerical evaluation of the
corresponding quantities, one finds the following scaling 
\begin{equation}
\frac{\xi _0^{\parallel }}{\xi _0^{\perp }}\simeq \frac{t_{\Vert }a_{\Vert }%
}{t_{\bot }a_{\bot }}.  \label{a10}
\end{equation}

To conclude this appendix let us stress the fact that expanding $\chi _0(%
{\bf q-Q}_{2D})$ to the second order using Eq. (\ref{a8}) to obtain the
asymptotic form of the $2D$-spin susceptibility is valid as long as the
maximum of $\chi _0$ is at $(\pi ,\pi )$, which is more general than
half-filling. Indeed, by symmetry, the first derivative of the free
susceptibility at ${\bf Q}_{2D}$ is zero for all fillings $n$ and
temperature and, as discussed before\cite{Schulz}\cite{Vilk}, at finite
temperature and away from half-filling the absolute maximum of the free
susceptibility can be at $(\pi ,\pi )$ even if it is not the case at $T=0$.
This behavior can be observed in Figure 5 where the in-plane second
derivative is plotted as a function of temperature for various values of
band filling. When the second derivative goes to zero there is a shift in
the wave vector maximizing the free susceptibility. Whether the magnetic
transition will be commensurate or incommensurate at a given filling depends
on the interaction $U$ because by changing $U$ one can change the ratio $\mu
/T_X$ and because the nature of the final three-dimensional order depends
very much at which wave vector correlations start to grow below $T_X$.\cite
{Vilk}

Calculations presented along the above lines do not allow us to study the
case where the maximum occurs at an incommensurate vector since we need the
analytical expressions for the second derivatives of $\chi _0$ to perform
very accurate numerical calculations (When the wave vector ${\bf q}$ is
different from $(\pi ,\pi )$ we do not have anymore the useful
simplification: $\epsilon _{{\bf k+q}}=-\epsilon _{{\bf k}}$ allowing us to
replace the $(k_x,k_y)$ integration in Eq.(\ref{a4}) by a simpler integral
on the $2D$ density of states.) Progress is nevertheless possible
numerically within the TPSC approach.

\section{Estimates for $\widetilde{\sigma }^2$ and $\Lambda ^{-1}$ in the
nearst-neighbor model.}

In this appendix we provide estimates for $\widetilde{\sigma }^2$ and $%
\Lambda ^{-1}$ in the isotropic two-dimensional nearest-neighbor model. The
surprisingly low numerical values $\tilde{\sigma}^2\simeq 0.21$, $\Lambda
^{-1}\simeq 0.022$ obtained for $U=4$ in the text are special to the Model C
perfect nesting case.

We first rearrange the self-consistency equation Eq.(\ref{sumSpin}) to
isolate the asymptotic behavior, as we did in Eq.(\ref{SelfQ2d}) but here in
two dimensions and with $a_{\bot }=1$.

\begin{equation}
n-2\langle n_{\uparrow }n_{\downarrow }\rangle =T\int \frac{d^2q}{\left(
2\pi \right) ^2}\chi _{sp}^{as}\left( {\bf q,}0\right) +T\int \frac{d^2q}{%
\left( 2\pi \right) ^2}\sum_{iq_n}\left[ \chi _{sp}\left( {\bf q,}%
iq_n\right) -\chi _{sp}^{as}\left( {\bf q,}0\right) \delta _{n,0}\right] .
\label{B1}
\end{equation}
It is usually assumed that the last integral on the right-hand side is
weakly temperature dependent and it is included with the left-hand side to
define $\widetilde{\sigma }^2$ . This procedure usually suffices for reasons
we will see below. For a more accurate estimate of $\widetilde{\sigma }^2$
close to $T_X$ we use the Euler-Maclaurin formula to approximate the sum
over Matsubara frequencies larger than the zeroth one by an integral.
Recalling also that $\chi _{sp}\left( {\bf q,}iq_n\right) =\chi _{sp}\left( 
{\bf q,-}iq_n\right) $ we have 
\begin{eqnarray}
n-2\langle n_{\uparrow }n_{\downarrow }\rangle &=&T\int \frac{d^2q}{\left(
2\pi \right) ^2}\chi _{sp}^{as}\left( {\bf q,}0\right) +T\int \frac{d^2q}{%
\left( 2\pi \right) ^2}\left[ \chi _{sp}\left( {\bf q,}0\right) -\chi
_{sp}^{as}\left( {\bf q,}0\right) \right]  \nonumber \\
&&+T\int \frac{d^2q}{\left( 2\pi \right) ^2}\chi _{sp}\left( {\bf q,}%
iq_1\right) +2\int_{2\pi T}^\infty \frac{d\lambda }{2\pi }\int \frac{d^2q}{%
\left( 2\pi \right) ^2}\chi _{sp}\left( {\bf q,}i\lambda \right) .
\end{eqnarray}

To recast this result in the same form as the consistency condition Eq.(\ref
{Consistency}), we first note that a more satisfactory definition of $%
\widetilde{\sigma }^2$ than the one given in Eq.(\ref{sigma2-2}) would be 
\begin{equation}
\widetilde{\sigma }^2=n-2\langle n_{\uparrow }n_{\downarrow }\rangle
-2\int_{2\pi T}^\infty \frac{d\lambda }{2\pi }\int \frac{d^2q}{\left( 2\pi
\right) ^2}\chi _{sp}\left( {\bf q,}i\lambda \right) .  \label{SigmaTilde}
\end{equation}
Also, the coefficient of the term linear in temperature on the right-hand
side of Eq.(\ref{Consistency}) would not only include the asymptotic
Lorentzian form but also a correction from the deviation to Lorentzian and
another correction from the first Matsubara frequency. Overall then, a more
accurate expression for the consistency condition is given by the last
definition of $\widetilde{\sigma }^2$ and 
\begin{equation}
\widetilde{\sigma }^2=T\left\{ \int \frac{d^2q}{\left( 2\pi \right) ^2}\chi
_{sp}^{as}\left( {\bf q,}0\right) +\int \frac{d^2q}{\left( 2\pi \right) ^2}%
\left[ \chi _{sp}\left( {\bf q,}0\right) -\chi _{sp}^{as}\left( {\bf q,}%
0\right) \right] +\int \frac{d^2q}{\left( 2\pi \right) ^2}\chi _{sp}\left( 
{\bf q,}iq_1\right) \right\} .  \label{Consistency2}
\end{equation}
The rest of this appendix is in two parts. We first estimate the left-hand
side of this equation, $\widetilde{\sigma }^2$ , and then we estimate the
integrals on the right-hand side to obtain $\Lambda ^{-1}$.

To obtain $\widetilde{\sigma }^2$, one should first notice that at the
crossover temperature the local moment $n-2\langle n_{\uparrow
}n_{\downarrow }\rangle $ is already quite close to its zero-temperature
value. Taking this as an estimate, we have 
\begin{equation}
n-2\langle n_{\uparrow }n_{\downarrow }\rangle =2\int_0^\infty \frac{%
d\lambda }{2\pi }\int \frac{d^2q}{\left( 2\pi \right) ^2}\chi _{sp}\left( 
{\bf q,}i\lambda \right)
\end{equation}
so that, substituting back into Eq.(\ref{SigmaTilde}), we have 
\begin{equation}
\widetilde{\sigma }^2=2\int_0^{2\pi T}\frac{d\lambda }{2\pi }\int \frac{d^2q%
}{\left( 2\pi \right) ^2}\chi _{sp}\left( {\bf q,}i\lambda \right) .
\end{equation}
To estimate this integral for the case of perfect nesting , we note that
singularities of $\chi _{sp}\left( {\bf q,}0\right) $ near wave vectors $%
{\bf q}=0$ and ${\bf q=}\left( \pi ,\pi \right) $ are integrable
singularities. We thus use the mean-value theorem to write, in our
dimensionless units 
\begin{equation}
\int \frac{d^2q}{\left( 2\pi \right) ^2}\chi _{sp}\left( {\bf q,}i\lambda
\right) \simeq \chi _{sp}\left( {\bf q}_{typ}{\bf ,}i\lambda \right)
\end{equation}
As a representative point, one can take ${\bf q}_{typ}=\left( \pi ,0\right) $
since it is far from both singularities. Using the trapezoidal rule to
estimate the frequency integral, one has 
\begin{equation}
\widetilde{\sigma }^2\simeq 2\frac{2\pi T}{2\pi }\left[ \frac{\chi
_{sp}\left( {\bf q}_{typ}{\bf ,}0\right) +\chi _{sp}\left( {\bf q}_{typ}{\bf %
,}2\pi T\right) }2\right] \simeq 0.19  \label{SigmaNum}
\end{equation}
whose numerical value follows from results obtained for $U=4$, $T_X\simeq
0.2 $, $U_{sp}\simeq U_{mf,c}\simeq 2$, 
\begin{eqnarray}
\chi _{sp}\left( {\bf q}_{typ}{\bf ,}0\right) &\simeq &0.60  \nonumber \\
\chi _{sp}\left( {\bf q}_{typ}{\bf ,}2\pi T_x\right) &\simeq &0.36.
\end{eqnarray}
The estimated numerical value of $\widetilde{\sigma }^2$ in Eq.(\ref
{SigmaNum}) corresponds closely to the value obtained in the text from
accurate numerical solutions. The fact that $\widetilde{\sigma }^2$ scales
roughly as $T_X\sim T_{mf,c}$ in very weak coupling, as follows from Eq.(\ref
{SigmaNum}), is a significant result since $\widetilde{\sigma }^2$ is also
related to the size of the pseudogap between precursors of antiferromagnetic
bands, as shown in Ref.\cite{Vilk2}.

To estimate the value of $\Lambda ^{-1}$, we notice that in the usual
consistency condition Eq.(\ref{Consistency}), one keeps only the first term
on the right-hand side of the more accurate expression Eq.(\ref{Consistency2}%
). The effect of the other terms is mimicked by using an effective cutoff $%
\Lambda $ that is not equal to the Brillouin zone size, as one might have
naively expected. In other words, the effective cutoff $\Lambda $ may be
obtained by requiring that 
\begin{equation}
\int_0^\Lambda \frac{qdq}{2\pi }\chi _{sp}^{as}\left( {\bf q,}0\right)
=\int_0^\pi \frac{qdq}{2\pi }\chi _{sp}^{as}\left( {\bf q,}0\right) +\int 
\frac{d^2q}{\left( 2\pi \right) ^2}\left[ \chi _{sp}\left( {\bf q,}0\right)
-\chi _{sp}^{as}\left( {\bf q,}0\right) \right] +\int \frac{d^2q}{\left(
2\pi \right) ^2}\chi _{sp}\left( {\bf q,}iq_1\right) .  \label{Cutoff}
\end{equation}
When there is no nesting, the quantity $\xi _0$ is relatively small at the
crossover temperature, meaning that the asymptotic Lorentzian form is not so
peaked and should be a good estimate of the susceptibility over much of the
Brillouin zone. Because of the slow decay of the asymptotic form $\chi
_{sp}^{as}\left( {\bf q,}0\right) $, the second integral should in fact be
negative and should partly cancel the last integral so that we should have $%
\Lambda \sim \pi $. By contrast, for perfect nesting $\xi _0\sim 1/T$ is
large, as seen in Appendix A, meaning that in this case the asymptotic form $%
\chi _{sp}^{as}\left( {\bf q,}0\right) $ is valid only in a narrow range of $%
{\bf q}$ values. Over most of the Brillouin zone, away from the maximum, the
true susceptibility $\chi _{sp}\left( {\bf q,}0\right) $ is larger than the
asymptotic one $\chi _{sp}^{as}\left( {\bf q,}0\right) $ because the latter
decays rapidly away from the maximum while the true one has an extremum at
both the Brillouin zone corner and center. The same arguments as those used
to evaluate integrals for $\widetilde{\sigma }^2$ allow us then to estimate 
\begin{eqnarray}
\int \frac{d^2q}{\left( 2\pi \right) ^2}\left[ \chi _{sp}\left( {\bf q,}%
0\right) -\chi _{sp}^{as}\left( {\bf q,}0\right) \right] &\simeq &\chi
_{sp}\left( {\bf q}_{typ}{\bf ,}0\right) \simeq 0.60 \\
\int \frac{d^2q}{\left( 2\pi \right) ^2}\chi _{sp}\left( {\bf q,}iq_1\right)
&\simeq &\chi _{sp}\left( {\bf q}_{typ}{\bf ,}2\pi T_x\right) \simeq 0.36
\end{eqnarray}
so that the equation Eq.(\ref{Cutoff}) that determines the cutoff becomes,
with $U_{sp}\simeq 2$ and $\xi _0^2\left( T_X\right) \simeq 1$, 
\begin{eqnarray}
\int_\pi ^\Lambda \frac{qdq}{2\pi }\chi _{sp}^{as}\left( {\bf q,}0\right)
&=&\int_\pi ^\Lambda \frac{qdq}{2\pi }\frac 2{U_{sp}\xi _0^2}\frac 1{\xi
^{-2}+{\bf q}^2}\simeq 0.96 \\
\Lambda ^{-1} &=&\pi ^{-1}\exp \left( -\frac{\pi U_{sp}\xi _0^2}20.96\right)
\simeq 0.016.
\end{eqnarray}
Although the difference with the numerically accurate result seems
relatively large, one should really compare the estimates of $\ln \Lambda
^{-1}$. The above estimate, $\ln 0.016\simeq -4.1$, differs only by roughly $%
10\%$ from the estimate, $\ln 0.022=-3.8$, obtained from a logarithmic plot
of the numerically accurate solution.

\section{Exact result for $\int {d^3q}\chi _{sp}^{as}$}

In this appendix we find the integral of the asymptotic part of the spin
susceptibility near ${\bf Q}_3=(\pi ,\pi ,\pi )$. Let $\chi _{as}({\bf q+Q}%
_3,0)$ be the approximate spin susceptibility near ${\bf Q}_3$ obtained in
Eq.(\ref{ChiAs}) with $q_{\perp }^2=q_x^2+q_y^2$ and $q_{\parallel }=q_z$
First we integrate in the $z$-direction from $-\Lambda _{\Vert }$ to $%
\Lambda _{\Vert }$, with $\Lambda _{\Vert }=\pi /a_{\Vert }$, then change to
polar coordinates in the plane and integrate on a circle of radius $\Lambda
_{\bot }$ to finally obtain 
\[
\int_D{\frac{d^3q}{(2\pi )^3}}\chi _{sp}^{as}({\bf q+Q}_3,0)={\frac 1{\pi
U_{sp}\left( \xi _0^{\bot }\right) ^2a_{\Vert }}}\left[ -{\frac 1{\Lambda
_{\Vert }\xi _{\Vert }}}{\rm arctg}\Lambda _{\Vert }\xi _{\Vert }\right. 
\]
\begin{equation}
\left. +{\frac 1{\Lambda _{\Vert }\xi _{\Vert }}}\sqrt{1+\Lambda _{\bot
}^2\xi _{\bot }^2}\ {\rm arctg}{\frac{\Lambda _{\Vert }\xi _{\Vert }}{\sqrt{%
1+\Lambda _{\bot }^2\xi _{\perp }^2}}}+{\frac 12}\ {\rm ln}(1+{\frac{\Lambda
_{\bot }^2\xi _{\perp }^2}{1+\Lambda _{\Vert }^2\xi _{\Vert }^2}})\right] .
\label{a2}
\end{equation}
This analytical result provides another route to obtain the N\'{e}el
temperature and the $d=2$ to $d=3$ crossover as discussed in the following
section.

\section{Extended scaling hypothesis and universality for the
renormalized-classical $d=2$ to $d=3$ crossover.}

We first briefly recall the results of Ref.\cite{PJF} on universality of
crossover scaling functions in anisotropic systems. The discussion usually
centers on anisotropy in spin space rather than position space but the
results are generally applicable. Suppose one has a very small anisotropy $g$%
. Sufficiently far from the transition, the critical behavior will be that
of the isotropic fixed point and should be described by the {\it extended
scaling hypothesis} for the singular part of the free energy density. The
same extended scaling hypothesis follows for other thermodynamic response
functions. We use the symbol $\left( \approx \right) $ to mean
''asymptotically equal to'' and $\left( \sim \right) $ to mean ''scales
as''. Let us concentrate on the magnetic susceptibility 
\begin{equation}
\chi \left( g,t\right) \approx At^{-\gamma }X\left( Bg/t^\phi \right)
\label{fsing}
\end{equation}
where 
\begin{equation}
t\equiv \left( \frac{T-T_c\left( 0\right) }{T_c\left( 0\right) }\right)
\end{equation}
with $\phi $ the crossover exponent and $T_c\left( 0\right) $ the value of
the transition temperature at zero anisotropy $g=0$. It is clearly the large
value of the correlation length that validates the scaling hypothesis. The
scale factors $A$ and $B$ in Eq.(\ref{fsing}) are non-universal, but the
scaling function is. The value of $A$ for a given model is fixed by the
normalization condition $X\left( 0\right) =1$.

Near the true transition temperature at the anisotropic fixed point, the
susceptibility should obey the usual result 
\begin{equation}
\chi \left( g,t\right) \approx \dot{A}\left( g\right) \stackrel{.}{t}^{-\dot{%
\gamma}}  \label{fani}
\end{equation}
where quantities with a dot refer to properties of the anisotropic fixed
point, and 
\begin{equation}
t\left( T,g\right) =\left( \frac{T-T_c\left( g\right) }{T_c\left( 0\right) }%
\right) +\left( \frac{T_c\left( g\right) -T_c\left( 0\right) }{T_c\left(
0\right) }\right) =\dot{t}+t_c\left( g\right) .
\end{equation}
The two expressions for the susceptibility Eqs.(\ref{fsing}) and (\ref{fani}%
) are consistent only if the crossover scaling function $X\left( Bg/t^\phi
\right) $ is singular as a function of its argument, namely 
\begin{equation}
\lim_{x\rightarrow x_c}X\left( x\right) =X_0\left( 1-\frac x{x_c}\right) ^{-%
\dot{\gamma}}  \label{ScalingFunction}
\end{equation}
where $X_0$ is a universal amplitude while $x_c$ is a $g$ and $t$
independent universal number. The definition 
\begin{equation}
x=Bg/\left( t\left( g\right) \right) ^\phi
\end{equation}
immediately implies that the transition temperature is at 
\begin{equation}
t_c\left( g\right) =\left( Bg/x_c\right) ^{1/\phi }.
\end{equation}

The generalization to the $d=2$ to $d=3$ crossover is not completely trivial
because in $d=2$ the correlation length is not a power law of temperature
for $O\left( n\right) $ models with $n>1$. Fisher and Barber\cite{Fisher} in
their study of crossover in helium films have considered the case where the
system is three-dimensional at high temperature and two-dimensional at low
temperature, opposite to the situation we consider. Furthermore, the
transition temperature is finite in $d=2$ helium films. Kosterlitz and Santos%
\cite{Kosterlitz} did consider the case of interest here, both within a
one-loop renormalization group approach, and in the spherical model. To cast
the results of the latter paper in the language of the extended scaling
hypothesis, it suffices to recall the usual hypothesis that the divergence
of the correlation length in the plane is at the origin of the scaling
behavior. Hence, $t\left( T\right) ^{-\nu }$ can be replaced everywhere in
the above equations by a function of absolute temperature that scales with $%
T $ in the same way as the two-dimensional correlation length\cite
{NotePuissance} 
\begin{equation}
\xi _{2d}\left( T\right) \equiv T^a\exp \left( C/T\right) .  \label{xsi2d}
\end{equation}
In this expression, we have allowed for a possible algebraic preexponential
factor. For example, to one loop order\cite{Kosterlitz} in the
momentum-shell method\cite{Pelcovits} the preexponential factor is $a=\left(
n-2\right) ^{-1}$ while to two-loop order\cite{Chakravarty88} as well as in
the $n\rightarrow \infty $ limit, only the exponential is present, $a=0$. In
addition to the non-universal quantities $A$ and $B$ defined above, we now
have an additional non-universal constant $C$ in Eq.(\ref{xsi2d}). This is
not fundamentally different from the usual case where the relation between $%
t $ and absolute temperature also involves a non-universal constant, namely $%
T_c\left( 0\right) $. The only difference between the itinerant case and the
usual $n-$vector model is that $C$ can be temperature dependent in the case
of nesting, as discussed in the text and in Appendix A. When there is no
nesting symmetry, $C$ is temperature independent. In the strong-coupling
limit, one usually defines $C=2\pi \rho _S$.

With the above $t\left( T\right) ^{-\nu }\rightarrow \xi _{2d}\left(
T\right) $ hypothesis, the extended scaling hypothesis becomes 
\begin{equation}  \label{GeneralSuscep}
\chi _{sp}^R({\bf Q}_d{\bf ,}0)\approx A\xi _{2d}^{\gamma /\nu }X\left(
Bg\xi _{2d}^{\phi /\nu }\right)
\end{equation}
where $g=\left( t_{\Vert }/t_{\bot }\right) ^2$ plays the role of the
anisotropy parameter in the case we have considered in detail in the text.
The function $X\left( x\right) $ is a universal function that we normalize
to $X\left( 0\right) =1$. With precisely the same asymptotic form as in Eq.(%
\ref{ScalingFunction}), simple power series expansion in powers of $T-T_N$
allows one to recover the correct critical behavior near the
three-dimensional N\'{e}el temperature. Hence, the N\'{e}el temperature is
given by $x_c=B\left( t_{\Vert }/t_{\bot }\right) ^2\xi _{2d}^{\phi /\nu }$
so that with the $n\rightarrow \infty $ result $\phi /\nu =\gamma /\nu =2$
and Eq.(\ref{xsi2d}) one recovers the result of the main text 
\begin{equation}  \label{Neel2}
\frac C{T_N}\sim \ln \left( \frac{t_{\bot }}{t_{\Vert }}\right) ^2
\end{equation}
with $C$ taking its appropriate temperature-dependent value in the nesting
case.

We conclude by an explicit calculation of the universal crossover function
for the staggered susceptibility in the $n\rightarrow \infty $ limit. In
this case, $a=0$ in Eq.(\ref{xsi2d}). The general form of the susceptibility
is given by Eq.(\ref{chiRPA}) with $\xi ^2=\xi _0^2\left( U_{sp}/\delta
U\right) $%
\begin{equation}
\chi _{sp}^R({\bf Q}_d{\bf ,}0)\approx \frac 2{\delta U}.  \label{chiQ}
\end{equation}
The value of $\delta U$ is in turn obtained by solving the self-consistency
condition Eq.(\ref{SelfQ2d}) 
\begin{equation}
\widetilde{\sigma }^2=Ta_{\Vert }a_{\bot }^2\int \frac{d^3q}{\left( 2\pi
\right) ^3}\chi _{sp}^{as}\left( {\bf q,}0\right)
\end{equation}
where $\widetilde{\sigma }^2$ takes essentially its $d=2$ value with very
small corrections. We can use the result of the previous appendix Eq.(\ref
{a2}) for the integral. It is then convenient to rewrite the result of the
integral in terms of the following dimensionless variables 
\begin{equation}
\alpha \equiv \frac{\Lambda _{\Vert }^2\xi _{\Vert }^2}{\Lambda _{\bot
}^2\xi _{\bot }^2}=\frac{\Lambda _{\Vert }^2\left( \xi _0^{\Vert }\right) ^2%
}{\Lambda _{\bot }^2\left( \xi _0^{\bot }\right) ^2}=cst\left( \frac{%
t_{\Vert }}{t_{\bot }}\right) ^2
\end{equation}

\begin{equation}
u\equiv \Lambda _{\Vert }\xi _{\Vert }=\Lambda _{\Vert }\xi _0^{\Vert
}\left( U_{sp}/\delta U\right) ^{-1/2}.  \label{Defu}
\end{equation}
Since we assume that we are in the scaling regime, namely the one where the
two-dimensional correlation length is very large, $\left( \Lambda _{\bot
}^2\xi _{\bot }^2\right) \gg 1$, we can use

\begin{equation}
\alpha \ll u^2
\end{equation}
to expand Eq.(\ref{a2}) and write 
\begin{equation}
\frac{\pi U_{sp}\left( \xi _0^{\bot }\right) ^2\widetilde{\sigma }^2}{%
Ta_{\bot }^2}=\ln \alpha ^{-1/2}+\ln \left( \frac{u^2}{u^2+1}\right)
^{1/2}+1-\frac 1u\arctan \left( u\right) .  \label{Implicit}
\end{equation}
If one solves the above implicit equation for $u$, then the susceptibility
Eq.(\ref{chiQ}) follows immediately from $u^2$ in Eq.(\ref{Defu}) since 
\begin{equation}
\chi _{sp}^R({\bf Q}_d{\bf ,}0)=\frac{2u^2}{\Lambda _{\Vert }^2\left( \xi
_0^{\Vert }\right) ^2U_{sp}}.  \label{SuscInt}
\end{equation}
Note that $u$ is a function of the dimensionless quantity $x$ defined by 
\begin{eqnarray}
x &\equiv &\alpha \xi _{2d}^2=\alpha \exp \left( 2C/T\right) =\alpha \Lambda
_{\bot }^2\xi _{\bot }^2 \\
C &\equiv &\frac{\pi U_{sp}\left( \xi _0^{\bot }\right) ^2\widetilde{\sigma }%
^2}{a_{\bot }^2}
\end{eqnarray}
as may be seen by exponentiating the implicit equation Eq.(\ref{Implicit}) 
\begin{equation}
x=\left( \frac{u^2}{1+u^2}\right) \exp \left( 2-\frac 2u\arctan u\right) .
\label{Implicit2}
\end{equation}

Before explicitly solving this equation in limiting cases, let us express
the universal scaling function $X$ in terms of $u$. The last equation for
the susceptibility Eq.(\ref{SuscInt}) may be rewritten with the above
definitions as 
\begin{equation}
\chi _{sp}^R({\bf Q}_d{\bf ,}0)=\frac{2u^2\left( x\right) }{\alpha \Lambda
_{\bot }^2\left( \xi _0^{\bot }\right) ^2U_{sp}}=A\xi _{2d}^2X\left(
x\right) =\frac{2\xi _{\bot }^2}{\left( \xi _0^{\bot }\right) ^2U_{sp}}%
X\left( x\right)
\end{equation}
where 
\begin{equation}
X\left( x\right) \equiv \frac{u^2\left( x\right) }x  \label{X(x)}
\end{equation}
and 
\begin{equation}
A\equiv \frac 2{\Lambda _{\bot }^2\left( \xi _0^{\bot }\right) ^2U_{sp}}.
\end{equation}

The universal crossover function $X\left( x\right) $ is plotted in Fig.6.
Let us check various limiting forms analytically. This will allow us to
recover all the cases studied in the main body of the paper. First, the
two-dimensional limit is the one where $u\rightarrow 0$. In this limit, the
implicit equation (\ref{Implicit2}) reduces to $x=u^2$. This verifies that
we have the proper normalization $X\left( x\right) =1$. The
three-dimensional limit is the limit where $u^2\rightarrow \infty $. In this
limit, 
\begin{equation}
x_c=e^2.
\end{equation}
Keeping the next term in the $1/u$ expansion, we have 
\begin{equation}
\lim_{x\rightarrow x_c}X\left( x\right) =\left( \frac \pi e\right) ^2\left(
1-\frac x{x_c}\right) ^{-2}
\end{equation}
hence the universal constant $X_0$ takes the value $\left( \pi /e\right) ^2$%
. As expected the susceptibility exponent in $d=3$ is $\dot{\gamma}=2$. The
N\'{e}el temperature follows from 
\begin{equation}
x_c=\alpha \exp \left( 2C/T_N\right)  \label{TNx}
\end{equation}
or 
\begin{equation}
\frac C{T_N}=\ln \left( \frac{e\Lambda _{\bot }\xi _{\bot }}{\Lambda _{\Vert
}\xi _{\Vert }}\right) \sim \ln \left( \frac{t_{\bot }}{t_{\Vert }}\right) .
\end{equation}
Finally, the $d=2$ to $d=3$ crossover temperature is given by $u=\Lambda
_{\Vert }\xi _{\Vert }=1$. Obviously, the criterion $\Lambda _{\Vert }\xi
_{\Vert }=1$ is subjective. We could take $\Lambda _{\Vert }\xi _{\Vert }$
to be equal to any other finite number. For definiteness however, we
continue with $\Lambda _{\Vert }\xi _{\Vert }=1.$ Substituting in Eq.(\ref
{Implicit2}) we have $x^{*}=\frac 12\exp \left( 2-\frac \pi 2\right) $ hence
the crossover temperature $T^{*}$ is given by 
\begin{equation}
x^{*}=\alpha \exp \left( 2C/T^{*}\right)
\end{equation}
Comparing with the equation for the N\'{e}el temperature Eq.(\ref{TNx}) we
find that the scaling of $\exp \left( C/T^{*}\right) $ with the anisotropy
parameter $\alpha $ is the same as that of $\exp \left( C/T_N\right) $. More
specifically, we find, in agreement with the main text, Eq.(\ref
{SizeCrossover}), that the size of the crossover region is given by, 
\begin{equation}
\left( 1-\frac{T_N}{T^{*}}\right) =\frac{T_N}{2C}\ln \left( x_c/x^{*}\right)
=\frac{\ln \left( x_c/x^{*}\right) }{\ln \xi _{2d}\left( T_N\right) }=\log
_{\xi _{2d}\left( T_N\right) }\left( x_c/x^{*}\right) .
\end{equation}
In other words, the size of the crossover region, calculated in reduced
temperature, decreases with $T_N$.

To complete the relation with the general functional form Eq.(\ref
{GeneralSuscep}) postulated above, note that if 
\begin{equation}
g\equiv \left( \frac{t_{\Vert }}{t_{\bot }}\right) ^2
\end{equation}
then $x=Bg\xi _{2d}^{\phi /\nu }=\alpha \xi _{2d}^2$ with $\phi /\nu =2$
implies that $B=\alpha /g$ is a number 
\begin{equation}
B\equiv \frac{\Lambda _{\Vert }^2\xi _{\Vert }^2}{\Lambda _{\bot }^2\xi
_{\bot }^2}\left( \frac{t_{\bot }}{t_{\Vert }}\right) ^2
\end{equation}
that is independent of $g$ because of the scaling $\Lambda _{\Vert }^2\xi
_{\Vert }^2/$ $\Lambda _{\bot }^2\xi _{\bot }^2\sim \left( t_{\Vert
}/t_{\bot }\right) ^2$ that follows from Appendix A, Eq.(\ref{a10}).

The universal crossover scaling function beyond $n=\infty $ where the
exponents $\phi /\nu \ $and $\gamma /\nu $ differ has yet to be investigated.

\begin{figure}
\centerline{\epsfxsize 6cm \epsffile{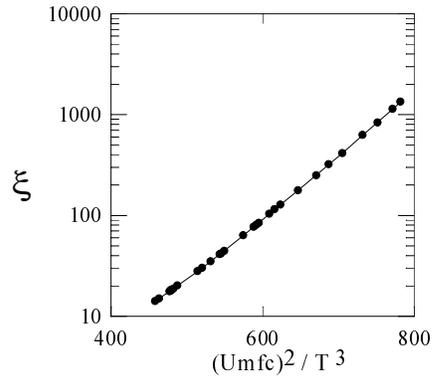}}
\caption{
Semi-logarithmic plot of the two-dimensional correlation
length, showing the scaling as a function of temperature in the case of
nesting.}
\end{figure}

\begin{figure}
\centerline{\epsfxsize 6cm \epsffile{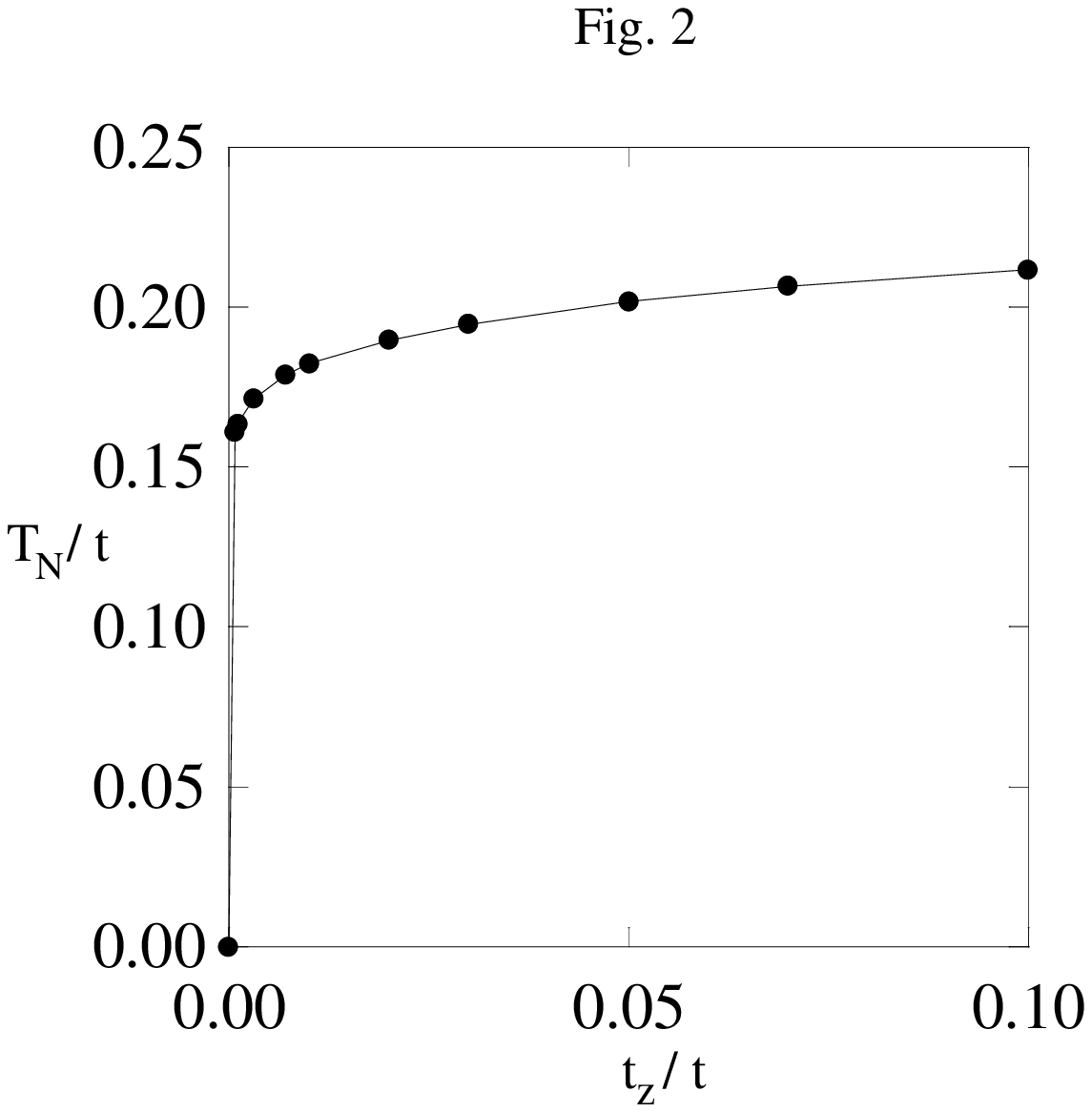}}
\caption{
N\'{e}el Temperature as a function of $t_z$ $\equiv
t_{\Vert }$ for $U=4t_{\bot }$, at half-filling.}
\end{figure}

\begin{figure}[tbp]
\centerline{\epsfxsize 6cm \epsffile{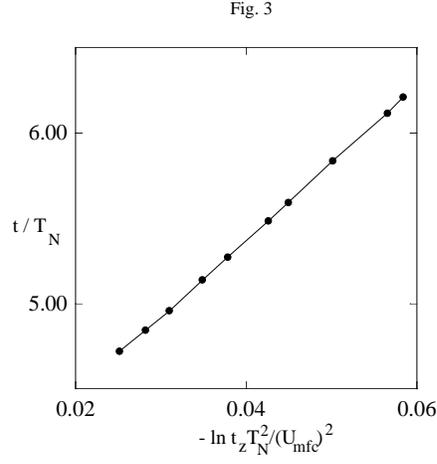}}
\caption{
${\frac 1{T_N}}$ as a function of ${\frac{T_N^2}{U_{mfc}^2}%
}\left| \ln {\frac{t_z}t}\right| $ for $U=4t_{\bot }$ at half-filling. The
quantities $T$ and $t_z$ $\equiv t_{\Vert }$ are in units of $t_{\bot }=t.$}
\end{figure}

\begin{figure}
\centerline{\epsfxsize 12cm \epsffile{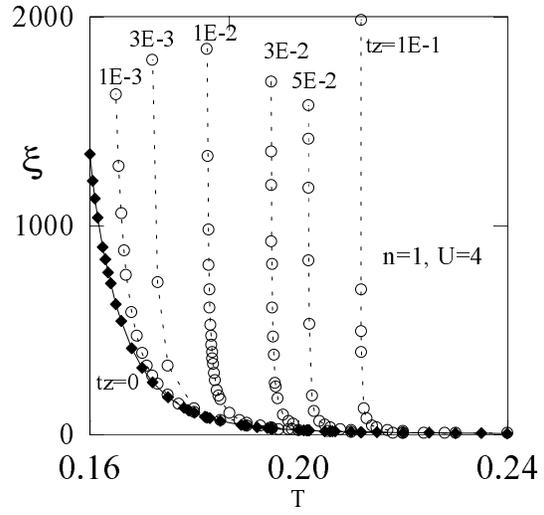}}
\caption{
In-plane correlation length $\xi _{\bot }$ (lattice space
is unity) for several values of out-of-plane hopping parameter at half
filling for $U=4t_{\bot }$. $T$ and $t_z$ $\equiv t_{\Vert }$ are in units
of $t_{\bot }$.}
\end{figure}

\begin{figure}
\centerline{\epsfxsize 10cm \epsffile{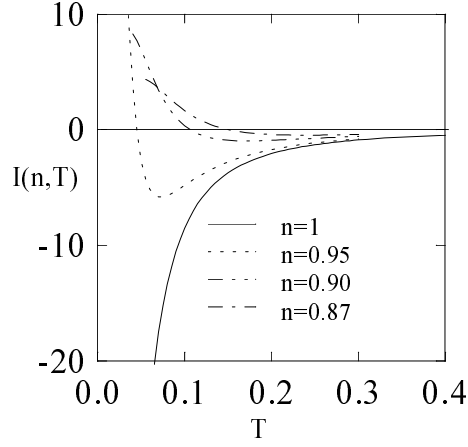}}
\caption{
Second in-plane derivative of the non-interacting susceptibility $%
I\left( n,T\right) =\partial ^2\chi _0\left( {\bf q,}0\right) /\partial
^2q_{\bot }|_{{\bf Q}}$ in two-dimensions as a function of temperature for
various band-fillings $n.$}
\end{figure}

\begin{figure}
\centerline{\epsfxsize 6cm \epsffile{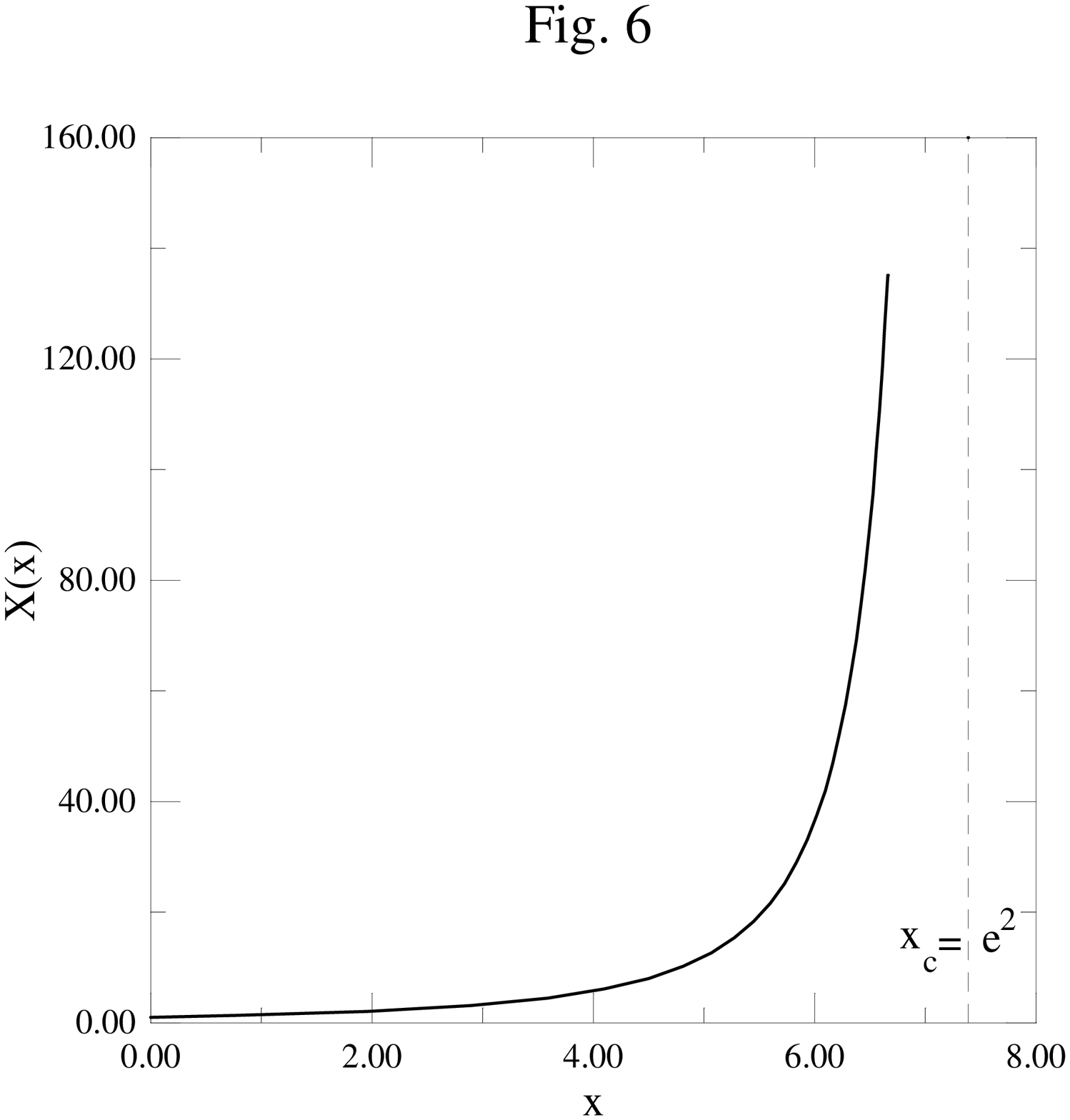}}
\caption{
Plot of the universal crossover function from $d=2$ to $d=3$ for
the staggered susceptibility in the $n\rightarrow \infty $ limit as defined
by Eqs.(\ref{X(x)}) and (\ref{Implicit2}).}
\end{figure}


\begin{references}
\bibitem{NoteT3}  In the practical situations discussed in the comparisons
with experiment, we have that $\xi $ varies between $1$ and $1500$ when the
temperature varies between $0.16t\ ${\rm and}$\ 0.2t$ so that the difference
between $1/T$ and $1/T^3$ scaling is not so important in the accessible
regime of temperatures.
\bibitem{NotePuissance}  In cases where the two-dimensional correlation
length has a power law prefactor, one should use $\xi _{2d}\left( T\right)
=\left( T/T_0\right) ^a\exp \left( C/T\right) $ where the value of $T_0$ is
determined by a normalization condition that we are free to choose. We can
choose $\xi _{2d}\left( T=1\right) =1$ for example. Once this normalization
is chosen, the non-universal quantities have a well defined value for any
given model.
\end{references}
\end{document}